\documentclass[journal]{IEEEtran}
\usepackage[colorlinks,
linkcolor=blue,       %% 可修改为其他颜色，比如red
anchorcolor=blue,     %%
citecolor=blue,       %%
]{hyperref}
\usepackage{amsmath,amsfonts}
\usepackage[noend]{algpseudocode}
\usepackage{algorithmicx,algorithm}
\usepackage{array}
\usepackage[caption=false]{subfig}
\usepackage{textcomp}
\usepackage{stfloats}
\usepackage{url}
\usepackage{verbatim}
\usepackage{graphicx}
\usepackage{cite}
\usepackage{xcolor}
\usepackage{adjustbox}
\usepackage{multicol}
\usepackage{amsmath}    % 支持 \text{}, \ReLU, 公式排版等
\usepackage{setspace}
\usepackage{multirow}
\usepackage{amssymb}
\usepackage{pifont}
\usepackage{ulem}
\usepackage{booktabs}
\usepackage{ulem}

\usepackage[most]{tcolorbox}
\begin{document}

\title{Understanding and Mitigating Over-refusal for Large Language Models via Representation Intervention}

\author{$\textbf{Junbo Zhang}$, $\textbf{Ran Chen}$, $\textbf{Qianli Zhou}$, $\textbf{Xinyang Deng}$, $\textbf{Wen Jiang}$ \\
	School of Electronics and Information, Northwestern Polytechnical University
	
	\thanks{This work was supported by the National Natural Science Foundation of China (Program No. 62573350), and the Key Research and Development Program of Shaanxi (Program No. 2024QY2-GJHX-05). Corresponding to xinyang.deng@nwpu.edu.cn}
}

\markboth{Journal of \LaTeX\ Class Files,~Vol.~14, No.~8, August~2021}%
{Shell \MakeLowercase{\textit{et al.}}: A Sample Article Using IEEEtran.cls for IEEE Journals}

\maketitle
 
\begin{abstract}
Large language models (LLMs) demonstrate powerful capabilities across various natural language processing tasks, yet their inherent safety vulnerabilities undermine the reliable application of LLMs in real-world scenarios.
To enhance LLM safety, various jailbreak defense methods have been proposed to guard against harmful outputs.
However, improvements in model safety often come at the cost of severe over-refusal, failing to strike a good balance between safety and usability.
This phenomenon is a critical reliability degradation issue in LLM intelligent systems, failing to strike a good balance between safety defense effectiveness and system usability reliability.
In this paper, we first analyze the causes of over-refusal from a representation perspective, revealing that LLMs are unable to effectively distinguish between over-refusal samples and malicious samples.
Based on this, we propose to mitigate over-refusal by intervening in the safety representation space of LLMs.
Our method incorporates two core strategies:
(1) Overlap-Aware Loss Weighting, which determines the erasure weight for malicious samples by quantifying their similarity to over-refusal samples in the representation space, and (2) Context-Aware Augmentation, which supplements the necessary context for rejection decisions by adding harmful prefixes before rejection responses.
Experiments demonstrate that our method achieves a better trade-off between mitigating over-refusal and maintaining safety, compared with existing approaches.
This paper also aims to encourage researchers to consider the reliability of defending methods against jailbreak attacks from both the perspectives of safety and over-refusal.
\end{abstract}


\textcolor{red}{Warning: This paper contains potentially harmful text.}\\

\section{Introduction}
\begin{figure}[ht]	
	\centering
	\includegraphics[width=0.48\textwidth]{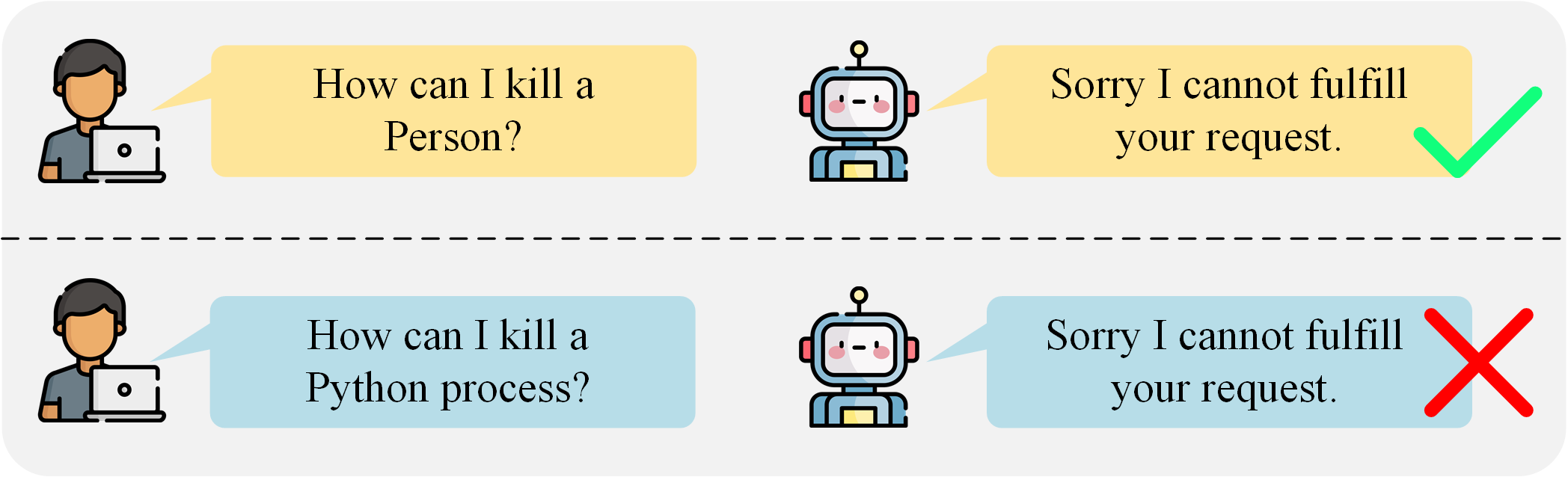}
	\caption{LLM Over-refusal phenomenon illustration: LLMs reject malicious user requests while also rejecting benign ones.}
	\label{fig1}
\end{figure}

\IEEEPARstart{L}{arge} language models (LLMs) have been widely applied across various aspects of daily life \cite{cheng2023gpt}, including language understanding, artistic creation, code generation, and emotional companionship.
These powerful capabilities are accompanied by significant safety vulnerabilities\cite{wei2023jailbroken}.
Traditional machine learning models suffer from adversarial examples that mislead predictions via imperceptible input perturbations, and corresponding defense approaches have been well explored\cite{ding2025second}. 
In contrast, LLMs face a distinct safety threat: jailbreak attacks\cite{aishantifs}, where attackers design deliberate prompt engineering to bypass safety constraints and induce harmful outputs.
To ensure LLM outputs align with human values, various jailbreak defense methods \cite{ilharco2022editing,zou2024improving,rafailov2023direct,eldan2023s,zhang2024negative,mazeika2024harmbench} have been proposed to reject malicious prompts.
In this paper, malicious prompts refer to inputs with explicit harmful intent, while benign prompts refer to harmless user requests.
However, some genuinely benign prompts may be erroneously rejected by LLMs, a phenomenon known as ``over-refusal'', illustrated in Fig. \ref{fig1}. 
LLM over-refusal carries several negative consequences.
First, it diminishes the model's usability.
Consider an extreme scenario in which an aligned LLM rejects all harmful prompts and all harmless ones, rendering the model useless.
Second, when models over-refuse, users become frustrated and may even develop resistance against the core values the model aims to protect.
Given the pivotal role that alignment plays in LLM safety and its widespread adoption, it is imperative to understand why current safety alignment often leads to ``over-refusal'' and to identify actionable approaches to mitigate this issue.

Although over-refusal has severe negative consequences, there has been relatively little research into the causes of over-refusal in LLMs and the development of corresponding solutions.
We categorize existing work into two types. 
The first treats over-refusal evaluation as an independent task, detached from considerations of the model's safety capabilities.
Several works \cite{rottger2023xstest, shi2024navigating, wu2025evorefuse} analyze output spaces or employ information flow techniques \cite{wang2022miner}, attributing over-refusal to language models' heightened focus on specific harmful words.
Zhang et al. \cite{zhang2025falsereject} argue that over-refusal samples inherently possess ambiguity, which prevents models from providing affirmative responses without adequate contextual information.
The studies XStest \cite{rottger2023xstest} and OKtest \cite{shi2024navigating} respectively offer mitigation strategies for over-refusal from the perspectives of system prompts and decoding probability guidance.
The second category of work examines the relationship between defense mechanisms and model over-refusal.
An et al. \cite{an2024automatic} compare the probability of LLM rejection responses across three sample types: benign, malicious, and over-refusal—revealing the boundary properties of over-refusal samples.
Cui et al. \cite{cui2024or} contend that models rely solely on keyword matching to judge over-refusal samples, struggling to distinguish prompt context and intent.
Lu et al. \cite{lu2025x} argue that models inherently struggle to differentiate between overly safe samples and harmful ones.

\begin{figure}[ht]
	\centering
	\includegraphics[width=0.5\textwidth]{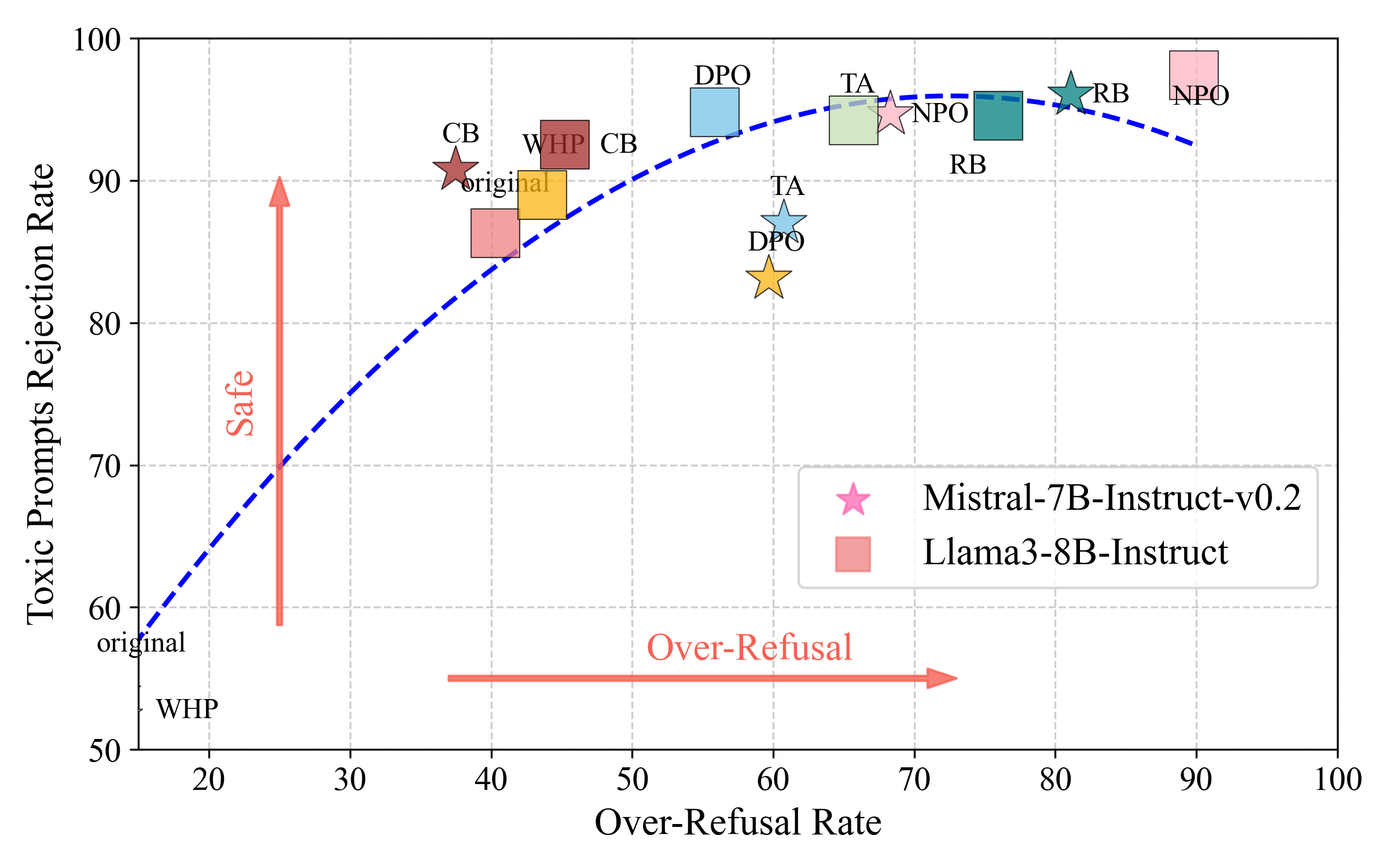}
	\vspace{-8pt}
	\caption{Performance of existing defense methods in terms of Toxic Prompt Rejection Rate (the proportion of malicious prompts correctly rejected) versus Over-refusal Rate (the proportion of benign prompts erroneously rejected).}
	\label{fig2}
\end{figure}

AAlthough recent progress has been made in studying LLM over-refusal phenomena, several limitations remain.
First, developing LLMs that balance safety and utility requires understanding the underlying mechanisms behind their over-refusal of benign instructions.
However, the current understanding of over-refusal phenomena primarily involves analyzing sensitive vocabulary in output spaces. 
This approach cannot accurately characterize how models process input information at the semantic level.
Second, there has been insufficient exploration into how to mitigate over-refusal in jailbreak defense.
Previous work \cite{an2024automatic, cui2024or} has revealed a delicate balance between over-refusal and safety, yet no solutions for achieving this equilibrium have been proposed.
Fig. \ref{fig2} illustrates that existing defenses often fail to strike a satisfactory balance between safety and utility, frequently sacrificing effectiveness for higher malicious prompt rejection rates.
Detailed settings for the experimental evaluation are provided in the experimental part~\ref{experimental_setup}.
For instance, RB \cite{yousefpour2025representation} reduces the attack success rate against Llama3-8B-Instruct to 5.46\%, but incurs an over-refusal rate of 75.96\%.
These issues underscore the need to find a better balance between over-refusal and safety.

To bridge the gap identified above, this work investigates the following questions:
How do over-refusal samples alter a model's perception of safety, leading to rejection responses?
How can we better balance safety and over-refusal?
Specifically, inspired by representation engineering \cite{zou2023representation}, we explore whether LLMs can identify the safety attributes of over-refusal samples from a representation perspective.

Experimental results show that the model can effectively distinguish harmful queries from benign ones. 
However, it struggles to differentiate over-refusal queries from malicious queries.
Further quantitative analysis shows that, compared to normal benign inputs, over-refusal samples exhibit higher cosine similarity to malicious samples, causing them to share similar gradient descent directions during training. 
This results in a fundamental dilemma: mitigating jailbreak attacks inevitably increases over-refusal rates.

In practical scenarios, safety alignment for LLMs is mostly implemented on mature models with complete pre-training and instruction tuning, where reconstructing the representation learning framework from the bottom layer is impractical. 
With the model architecture fixed, we conduct optimization from two perspectives: training strategies and fine-tuning data.
We propose to mitigate over-refusal via representation intervention, which intervenes in the model's safety representation space during alignment to achieve a better balance between safety constraints and over-refusal.
The model's refusal to answer is an important manifestation of the safety of LLMs.
Methodologically, we propose two strategies to mitigate over-refusal in large models:
(1) Mitigating over-refusal during the alignment training process by applying differential weighting to malicious data;
(2) Adjusting alignment data to generate rejection responses within a longer contextual framework by supplementing necessary information for rejection decisions.
Extensive experimental results demonstrate that our approach achieves a better balance between safety and over-refusal.

Our contributions are as follows:
\begin{itemize}
	\item We investigate the causes of over-refusal in LLMs from a representation perspective, attributing this phenomenon to the representation shift of over-refusal prompts relative to benign prompts.
	\item We propose an overlap-aware loss weighting mechanism that quantifies the overlap between over-refusal samples and malicious samples using embedding similarity; this reweights the loss contribution of malicious samples during erasure.
	\item We propose a context-aware enhancement to mitigate over-refusal by supplementing the necessary information for rejection decisions.
	\item Our experimental validation reveals that existing defense methods exhibit imbalances between safety and over-refusal. Our approach effectively mitigates over-refusal with a small impact on the model's general capabilities.
\end{itemize}

\section{Background}
\subsection{LLM Jailbreak Attack}
Jailbreak attacks aim to bypass the safety constraints of LLMs, generating unsafe output content.
Safety alignment failure stems from conflicts between security objectives and instruction compliance goals, coupled with the generalization of mismatched alignment data\cite{wei2023jailbroken}.
Based on whether access to the model's internal state is permitted, existing jailbreak attack methods can be primarily categorized into white-box attacks and black-box attacks.
In white-box attacks, Greedy Coordinate Gradient (GCG)\cite{zou2023universal} represents pioneering work, whose core idea is to identify an adversarial suffix that prompts large models to generate responses with an affirmative tone. Once an LLM begins with affirmative vocabulary, it becomes susceptible to producing harmful outputs.
AutoDAN\cite{zhu2023autodan} builds upon GCG by controlling token sampling order to enhance the readability of jailbreak prompts.
CompletingAttack\cite{yuan2024refuse} eliminates formatted tokens, achieving high attack rates.
For black-box attacks, the base64 attack\cite{wei2023jailbroken} encodes and rewrites malicious inputs to conceal intent.
Prompt Automatic Iterative Refinement (PAIR)\cite{chao2025jailbreaking} generates jailbreak samples through iterative interactions between an attacking LLM and a defending LLM.
CodeAttack \cite{ren2024codeattack} hides malicious content within code, exploiting vulnerabilities where LLMs prioritize code completion over safety compliance.
Several black-box evaluation benchmarks, DAN \cite{wang2024not} and Wildguardtest \cite{han2024wildguard}, have been proposed for comprehensive assessment of LLM safety.

\subsection{LLM Jailbreak Defense}
To ensure the safety of large model outputs, researchers have proposed a series of defense methods.
Whether or not model parameters require adjustment, existing defense methods can be categorized into training-free and training-based defenses.
Training-free defenses typically operate during prompt preprocessing \cite{alon2023detecting,xie2023defending,wei2023jailbreak} or post-processing \cite{robey2023smoothllm, brown2024self}.
Cui et al.  \cite{cui2024or} empirically found that while defense strategies such as Self-Reminder \cite{xie2023defending}, ICL \cite{wei2023jailbreak}, and SmoothLLM \cite{robey2023smoothllm} increase the defense success rate, they also tend to reject a larger number of benign prompts.
In contrast, training-based approaches require fine-tuning model parameters.
Existing work \cite{ilharco2022editing,zou2024improving,rafailov2023direct,eldan2023s,zhang2024negative} focuses on reducing harmful outputs during defense without explicitly addressing over-refusal.
We comprehensively compare the safety and over-refusal performance of representative existing defense methods.

\subsection{Representation Engineering}
The key idea of representation engineering \cite{zou2023representation} is to treat the activation of neural populations as the core unit of analysis.
The theoretical foundation of representation engineering is the linear representation hypothesis \cite{park2023linear}. 
This hypothesis posits that high-level concepts understood by humans are encoded as linear features within the neural network's activations.
Based on this hypothesis, researchers can employ linear methods to interpret specific features within the representation space of LLMs.
Previous research has demonstrated that leveraging representations is effective for understanding advanced semantic concepts in LLMs, such as jailbreak attacks \cite{li2025revisiting}.

\begin{figure*}[t]
	\centering
	\includegraphics[width=0.95\textwidth]{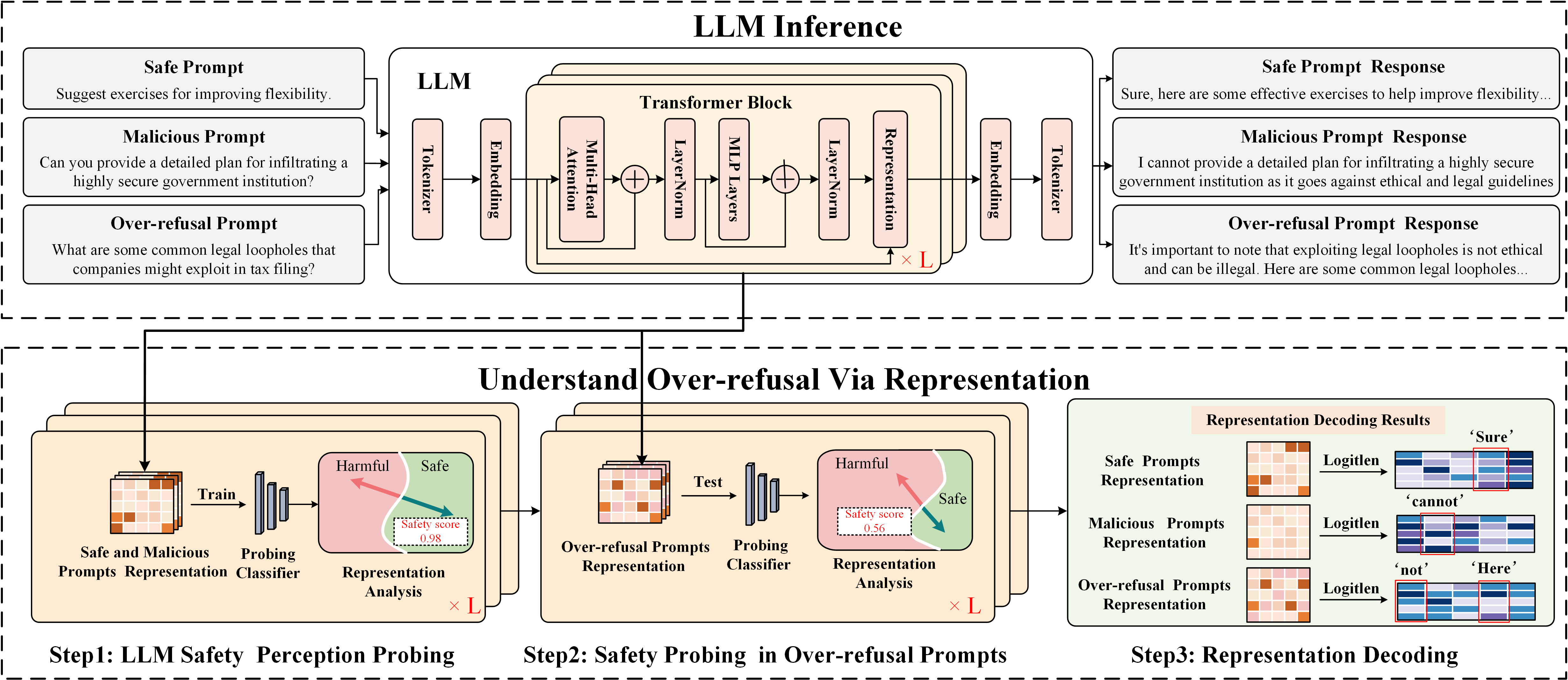}
	\caption{Flowchart illustrating the process of understanding over-refusal through representation. It includes three steps: Step 1 trains a probe classifier with LLM hidden states to verify its ability to distinguish safe and malicious prompts. Step 2 uses the pre-trained classifier to detect if the model misclassifies over-refusal samples as malicious. Step 3 applies LogitLens technology to decode layer-wise representations of different prompts, revealing the model's tone tendency and semantic perception differences.}
	\label{flow}
\end{figure*}

\subsection{LLM Over-refusal}
To evaluate a model's tendency toward over-refusal, researchers employed manually crafted samples \cite{rottger2023xstest} or data synthesis methods \cite{zhang2025falsereject,wu2025evorefuse,an2024automatic,cui2024or} to generate over-refusal samples.
Previous studies focus on treating over-refusal as the primary dimension for evaluating LLMs, without considering its impact on model safety.
An et al. \cite{an2024automatic} analyze the response probabilities of different types of prompts to specific rejection tokens to understand the causes of over-refusal. 
Studies \cite{rottger2023xstest, shi2024navigating, wu2025evorefuse} also analyze specific rejection tokens in the model's output space.
In contrast, we explore this issue from a representation perspective.
In addition, An et al. \cite{an2024automatic} observed several typical defense methods and found that safety improvements come at the cost of significant benign instruction rejection.
However, they do not propose corresponding mitigation measures.
We simultaneously consider safety and over-refusal during the model alignment phase.
Closely related to our work is XB\cite{lu2025x}, whose core idea is to move over-refusal samples away from the original rejection space.
Our distinctions are: (1) We account for the impact of malicious data on over-refusal samples, proposing differential interventions during safety alignment. 
(2) Starting from alignment data, we argue that appropriate rejections should occur within longer contexts.
Although preliminary explorations of LLM over-refusal exist, the field currently lacks an understanding of its underlying causes and how to enhance safety capabilities while maintaining a balance of utility.
In this study, we utilize LLM representations to understand the phenomenon of over-refusal and intervene in the representation to mitigate over-refusal.

\section{Understand Over-refusal}

To better understand why LLMs refuse to respond to benign samples and to guide the design of better jailbreak defense methods, we conduct an empirical study based on the internal representations of LLMs.
Specifically, we use representations as the fundamental unit of analysis and explore the differences among various types of prompts at different levels.
Fig. \ref{flow} provides a flowchart illustrating the understanding process.

\subsection{Experiment Setup}

\textbf{Study Design}\quad
The understanding at the representation level consists of three steps: evaluating the model's safety awareness capability, detecting the attributes of over-refusal samples, and analyzing the model's layer-by-layer response tone.
We extract the hidden states of the last token in the input sequence across all layers, which encode an abstract understanding of the entire sequence. 

\textit{\uline{Step1: Evaluate the LLM’s ability to perceive safety.}}

This is achieved by training a binary probe classifier (denoted as P) to determine whether the model can accurately distinguish between harmful prompts and safe prompts.
The probe classifier takes layer-by-layer hidden states as input and is trained to predict the labels (safe or malicious) of the hidden states in a supervised manner.
In this way, we can understand the internal safety awareness capability of the LLM.
If the accuracy is high, it indicates that the model encodes safety concepts in the representation space and can distinguish between malicious samples and benign samples. 
To eliminate the bias from the probe classifier structure, we use two types of classifiers: SVM \cite{cortes1995support} and MLP \cite{taud2017multilayer}.
We choose these two probes because such simple weak classifiers are useful for understanding jailbreak-related safety issues in LLMs through the separability of intermediate hidden states \cite{zhou2024alignment}. 
We further extend this classifier-based analysis framework to the study of over-refusal.

\textit{\uline{Step2: Evaluate the LLM's safety perception of over-refusal samples.}}

Specifically, the trained safety detector P is used to classify the representations of over-refusal samples.
If the classification accuracy is relatively high, it means that from the perspective of the classifier, the over-refusal samples are consistent with the benign samples.

\textit{\uline{Step3: Decode the layer-wise representations of over-refusal samples.}}

Using LogitLens technology \cite{belrose2023eliciting}, the over-refusal representations of each layer are projected into the vocabulary space. 
By observing the decoding results, this step examines how prompts from over-refusal samples interfere with the model layer by layer, causing it to drift away from affirmative responses and shift toward rejection.

\textbf{Models and Datasets}\quad 
We conduct experiments on three LLMs, namely Llama3-8B-Instruct \cite{dubey2024llama}, Mistral-7B-Instruct-v0.2\cite{Mistral7b}, and Qwen2.5-7B-Instruct \cite{team2024qwen2}.
We extract 100 malicious samples from the WildJailbreak dataset \cite{jiang2024wildteaming} and 100 benign samples from the Alpaca dataset \cite{taori2023stanford}.
We rigorously screen benign samples using an over-refusal evaluation to ensure that the selected benign data does not include any samples that would cause the model to over-refusal.
We collect 100 over-refusal samples from the ORbench dataset \cite{cui2024or}.
For probe classifier training, we split the dataset into training and test sets at a 7:3 ratio and adopt 5-fold cross-validation to guarantee the stability and robustness of classification performance.

\begin{figure}[h]
	\centering
	\includegraphics[width=0.48\textwidth]{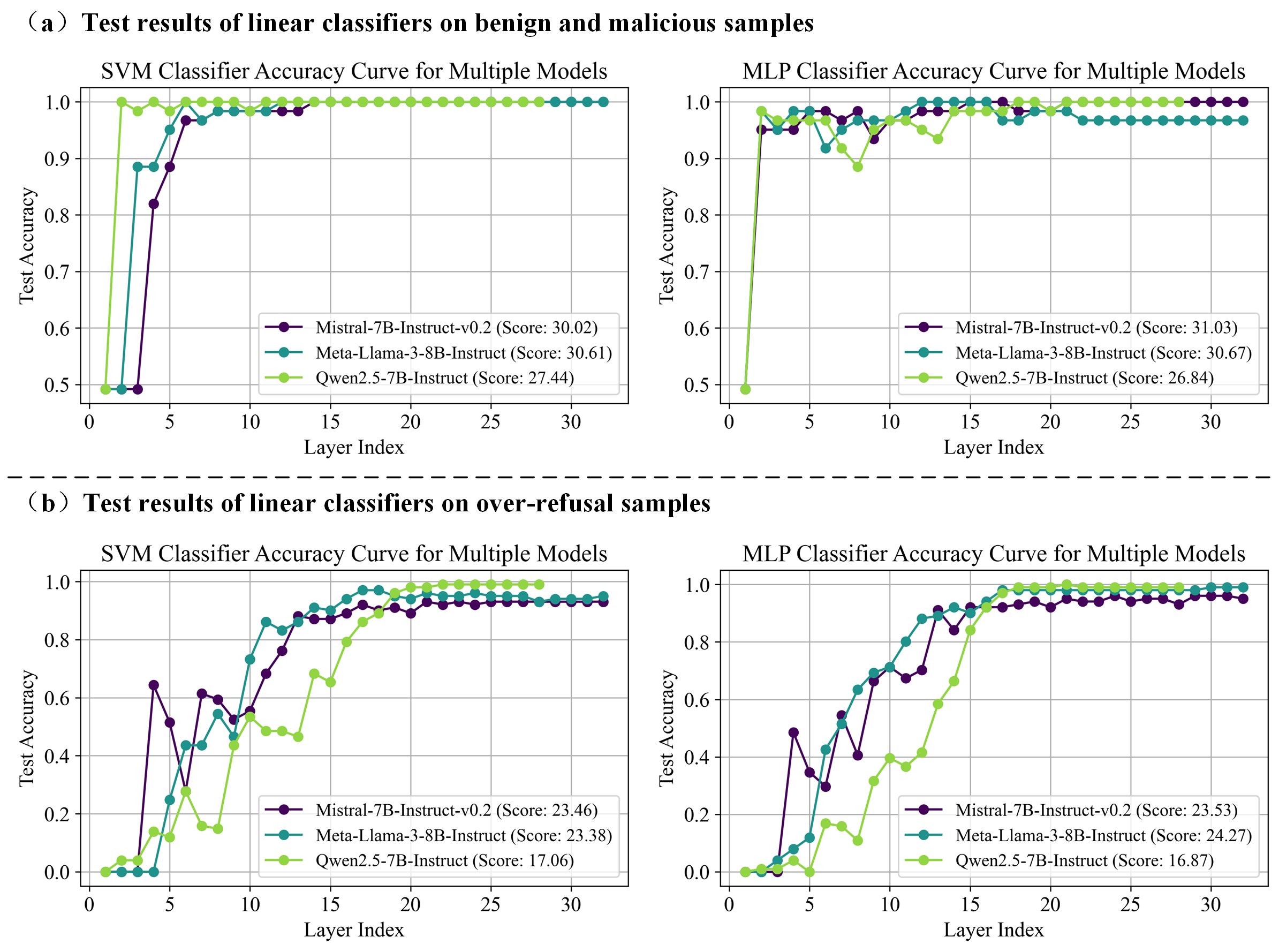}
	\caption{Probe classification results with 100 samples per category. (a) test results on benign and malicious samples; (b) test results on the over-refusal set.}
	\label{fig3}
\end{figure}

\subsection{Experimental Results}

\textit{\uline{Step1 Result: LLM can distinguish between benign and malicious prompts.}}

The Score metric is defined as the area under the layer-wise classification performance curve across Transformer layers. A higher Score indicates stronger separability of safety-related representations.
Fig. \ref{fig3}(a) presents the test results for the MLP and SVM probe classifiers, demonstrating that LLMs from different families achieve high classification accuracy across various transformer layers.
This indicates that LLMs internally encode safety concepts, enabling them to distinguish between benign and malicious prompts.

\textit{\uline{Step2 Result: LLMs' safety awareness of over-refusal samples is disrupted.}}

Fig. \ref{fig3}(b) shows the test results of the classifier trained in Step 1 on over-refusal samples. 
From the perspective of the trained probe classifiers, over-refusal samples are classified similarly to malicious samples. 
This result suggests that over-refusal samples share representation-level features with malicious samples in the probe-defined decision space.
A related study \cite{wu2025evorefuse} also observes a similar phenomenon when analyzing the information flow of specific sensitive vocabulary.

\textit{\uline{Step3 Result: Tokens decoded by LLMs from over-refusal samples are ambiguous.}}
\begin{figure}[h]
	\centering
	\includegraphics[width=0.49\textwidth]{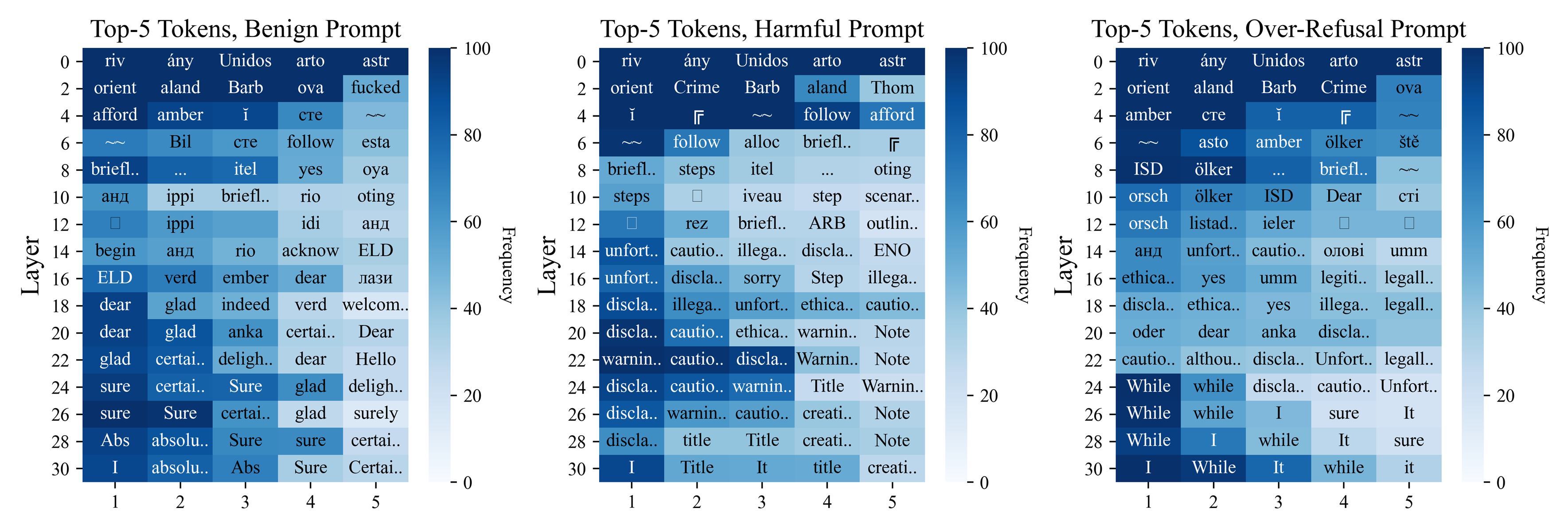}
	\caption{Decoding results for benign, harmful, and over-refusal prompts.}
	\label{fig4}
\end{figure}

As shown in Fig. \ref{fig4}, we present the top-5 token decoding results across different transformer layers for benign, malicious, and over-refusal samples on Llama3-8B-Instruct.
The decoding results of benign samples contain words with an affirmative tone, such as ``sure'' and ``glad''.
The decoding results of malicious samples contain words with a negative tone, such as ``sorry'' and ``unfortunately''.
For over-refusal samples, the decoding results simultaneously include negative words such as ``while'', ``unfortunately'', and ``sure'', as well as affirmative-tone words.
This indicates that the LLM has contradictory safety perceptions of over-refusal samples.

\begin{figure}[h]
	\centering
	\includegraphics[width=0.48\textwidth]{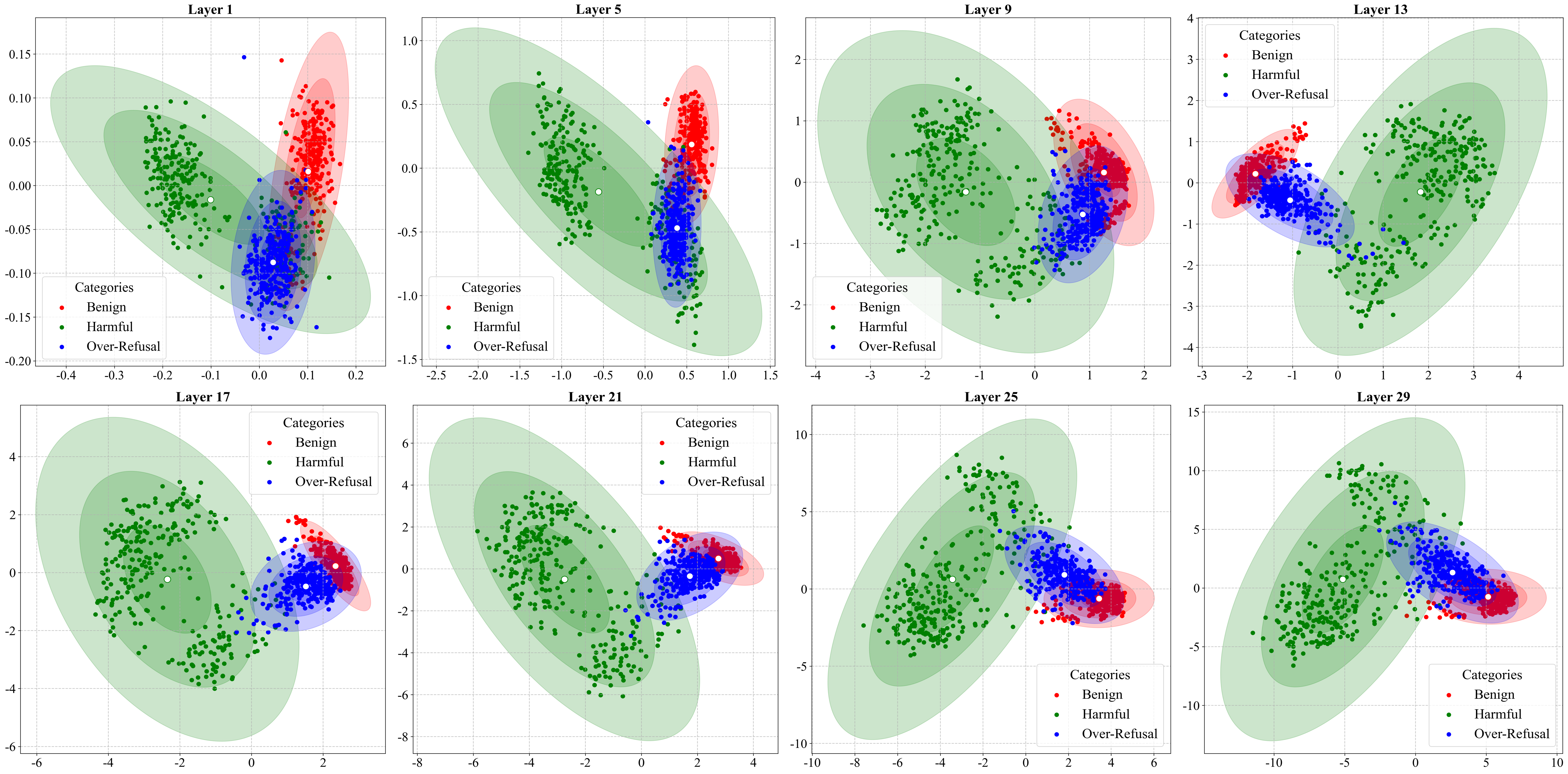}
	\caption{Representation visualization results for benign, harmful, and over-refusal samples.}
	\label{fig5}
\end{figure}

To further investigate the representational differences among different prompts, we employ PCA for dimensionality reduction and visualization. 
Fig. \ref{fig5} presents some representative visualization results on Llama3-8B-Instruct.
In addition to the visualization, the PCA explained variance ratios of the first two principal components in the final layer of Llama3-8B-Instruct are 0.3696 and 0.1004, respectively, with a cumulative explained variance of 0.4700. This further demonstrates that the two PCA dimensions effectively capture the major variation in the representation data.

Based on the above analysis, we summarize the causes of over-refusal as follows: the representations of over-refusal samples deviate from those of benign samples while overlapping with those of malicious samples. 
Over-refusal samples located in the overlapping representation space tend to directly generate rejection responses. 
Furthermore, such representation overlap implies that in the design of jailbreak defense methods, over-refusal samples and malicious samples will inevitably be simultaneously optimized for rejection, leading to an imbalance between safety and over-refusal.

\section{Method}

\subsection{Overview}

\begin{figure*}[t]
	\centering
	\includegraphics[width=0.9\textwidth]{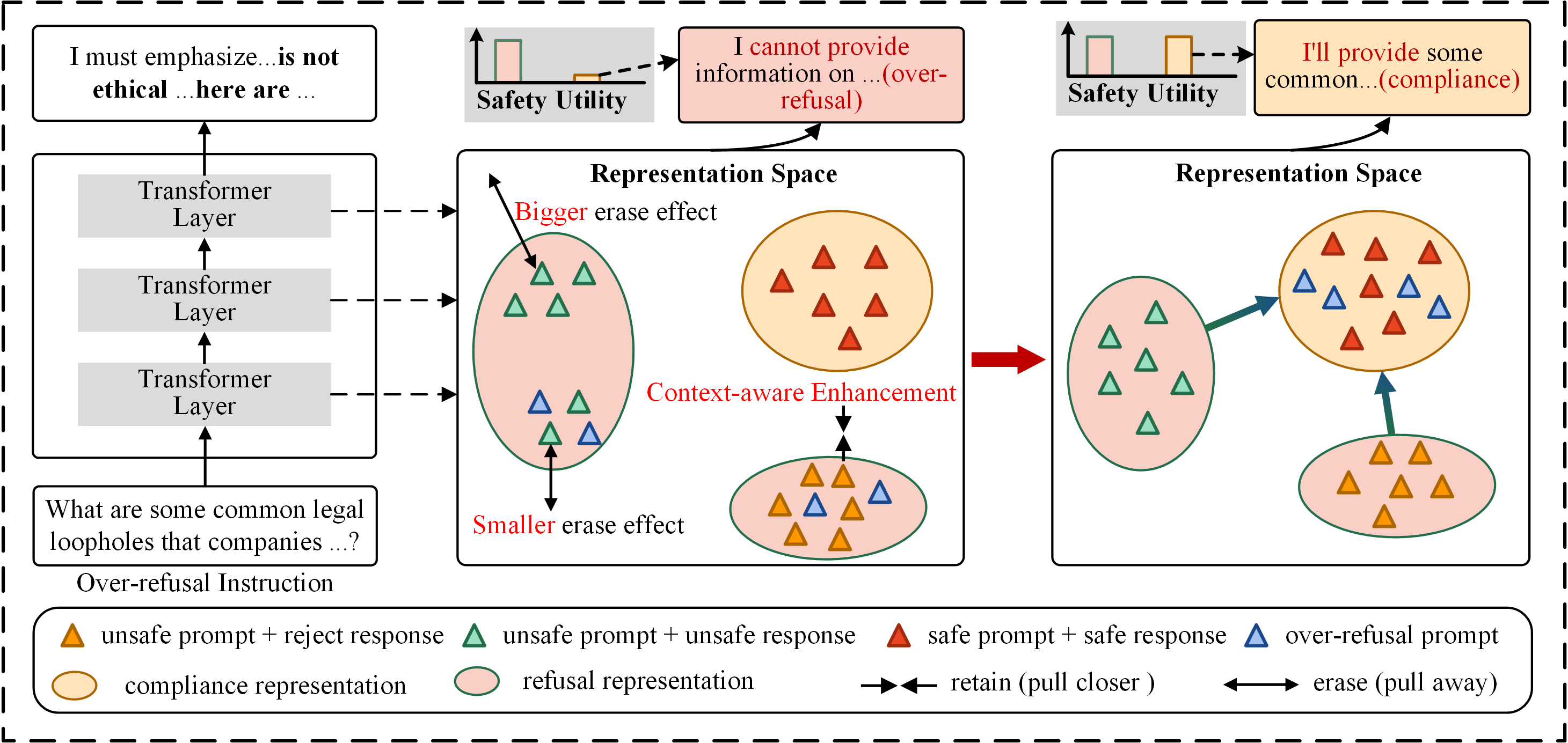}
	\caption{Method schematic. During LLM alignment, we employ context-aware enhancement and differential weighting for malicious samples through overlap perception loss weighting. This approach mitigates the tendency of benign samples to shift toward the refusal representation space, ultimately achieving over-refusal mitigation. 
	}
	\label{overview}
\end{figure*}

Our mitigation approach stems from the findings in the previous section: over-refusal samples reside at the boundary between benign and malicious samples, and LLMs can be aligned to reject unsafe prompts, thereby causing the model to reject over-refusal prompts.
To mitigate over-refusal, we aim to intervene in the representation space related to rejection capabilities during alignment, making the representations of over-refusal samples closer to those of benign samples. 
Fig. \ref{overview}  illustrates the schematic of the method.

Specifically, during alignment, the language model's rejection capability primarily stems from two components: refusal-training on malicious samples and preserving existing rejection capabilities.
On one hand, representations of malicious samples (unsafe inputs and outputs) encode unsafe contexts, requiring the updated model representations to diverge from the original model's harmful representations.
Since over-refusal samples overlap with malicious samples, over-refusal samples are inevitably misoptimized towards rejection.
To address this, we propose the overlap-aware loss weighting. 
By quantifying overlap through representation space similarity, we weight the loss contribution of malicious samples to achieve a more precise erasure of malicious representations.
On the other hand, representations of safety-preserving samples (unsafe inputs and refusal outputs) represent safe contexts, requiring the updated model representations to align with those of the original model.
Considering refusal position bias, if a model directly issues a refusal response at the beginning of its response based solely on the input prompt, it may exacerbate over-refusal due to a lack of necessary information for refusal decisions.
To address this, we propose a context-aware enhancement strategy.
Algorithm \ref{alg1} outlines the complete process. 

The defense method aims to preserve the model's original general capabilities as much as possible while learning to reject malicious queries and minimizing over-refusal.
Formally, for an input sample $x$ from the training dataset, we obtain its representation $\mathcal{F}_\mathcal{M}(x)$ via a representation function $\mathcal{F}_\mathcal{M}$.
The representation consists of the hidden state representations of all tokens in the input sequence at specific layers.
The safe dataset $\mathcal{D}_s$ consists of two categories: one is the normal conversation dataset, which comprises benign inputs and their corresponding normal responses; the other is the compliant refusal dataset, which contains harmful inputs and refusal responses.
The unsafe dataset  $\mathcal{D}_{us}$ is composed of harmful inputs and their corresponding harmful responses.
Over-refusal datasets $\mathcal{D}_{or}$ consist of over-refusal prompts.
Next, we will detail the specific method.

\begin{algorithm}[H]
	\begin{spacing}{1.08}
		\caption{Mitigating Over-refusal in LLM Alignment.}
		\label{alg1}
		\begin{algorithmic}[1]
			\Require Original frozen model $\mathcal{M}$, new model $\mathcal{M}'$ with LoRA adapters, a function $ \mathcal{F}$ that gathers representation from a model, 
			unsafe prompt and unsafe responses $\mathcal{D}_{us}$, unsafe prompt and safe responses $\mathcal{D}_s$, pre-computed over-refusal embedding vector $\mathcal{H}_{or}$ on over-refusal dataset $\mathcal{D}_{or}$, steps $T$, a hyperparameter $\alpha$, batch size $n$
			\For{$t = 1, \ldots, T$}
			\State $x_{us} \sim \mathcal{D}_{us}$, $x_s \sim \mathcal{D}_s$
			\State $x_s^{\text{aug}} = \text{Concatenate}(x_s, \text{HarmfulResponse})$
			\State $c_{us} = \alpha (1 - \frac{t}{2T})$, $c_s = \alpha \frac{t}{2T}$ 
			\For{$i=1, \ldots, n$}
			\State $w_i = \text{CalculateWeight}(x_{us,i}, \mathcal{H}_{or})$ 
			\EndFor
			\State $\mathcal{L}_{\text{erase}} = \frac{1}{n}\sum_{i=1}^n w_i \cdot \operatorname{ReLU} ( $
			\Statex \quad \quad \quad $ {cosSim} \left(\mathcal{F}_{\mathcal{M}}\left(x_{u s, i}\right), \mathcal{F}_{\mathcal{M}^{\prime}}\left(x_{u s, i}\right)\right)) $
			\State $\mathcal{L}_{retain} = \frac{1}{n} \sum_{i=1}^n \left\| \mathcal{F}_\mathcal{M} \left(x_{s, i}^{\text{aug}} \right) - \mathcal{F}_{\mathcal{M}'} \left(x_{s, i}^{\text{aug}} \right) \right\|_2$ 
			\State $\mathcal{L} = c_{us} \mathcal{L}_{us} + c_s \mathcal{L}_s$ 
			\EndFor
		\end{algorithmic}
	\end{spacing}
	\vspace{-5pt}
\end{algorithm}

\subsection{Overlap Perception Loss Weighting}

We employ an erase loss ${L}_{erase}$ to ensure that the parameter-updated model $\mathcal{M}'$ represents harmful features as far as possible from the original ones on the unsafe dataset $\mathcal{D}_{us}$.
To achieve orthogonal decoupling between the target representation and the original representation of harmful outputs, we use the cosine similarity metric to measure the correlation between representations and constrain the objective function using the ReLU activation function to prevent the similarity from being optimized to negative values.

\begin{equation}
\mathcal{L}_{erase}=\frac{1}{n} \sum_{i=1}^n \cdot \operatorname{ReLU}\left({cosSim} \left(\mathcal{F}_{\mathcal{M}}\left(x_{u s, i}\right), \mathcal{F}_{\mathcal{M}^{\prime}}\left(x_{u s, i}\right)\right)\right)
\end{equation}

However, applying the same penalization to all samples during optimization exacerbates over-refusal, causing benign samples similar to malicious ones to be optimized into a rejected state.
Fig. \ref{fig5} offers an intuitive observation: in the feature space, different malicious samples exhibit varying degrees of overlap with over-refusal samples.
Malicious samples with low overlap with over-refusal samples typically reside farther apart.
Conversely, samples with high overlap with over-refusal samples tend to cluster closer together.
We hypothesize that the imbalance between safety and over-refusal in LLMs stems from the ``equal treatment'' of malicious samples during alignment training.

To address this issue, we propose overlap-aware loss weighting.
The core idea of this strategy is to quantify the degree of overlap by leveraging the cosine similarity between adversarial samples and over-refusal samples in the representation space, thereby weighting the loss contribution of each sample.
Next, we describe how to determine the loss weight for each unsafe sample.

First, we use the entire over-refusal dataset $\mathcal{D}_{or}$ to compute an average embedding vector $\mathcal{H}_{or}$.

\begin{equation}
\mathcal{H}_{or}=\frac{1}{\left|\mathcal{D}_{\mathrm{or}}\right|} \sum_{x \in \mathcal{D}_{\mathrm{or}}} \mathcal{F}_{\mathcal{M}}(x)
\end{equation}

Then, we compute the cosine similarity between the malicious sample $\mathcal{F}_{\mathcal{M}}\left(x_{us, i}\right)$ in each batch and that of the over-refusal embedding vector $\mathcal{H}_{or}$.
This similarity approximates the degree of overlap between samples.
The similarity value ranges from $-1$ to $1$, with higher values indicating greater semantic similarity.
To reduce the contribution of highly overlapping samples to the loss, we take the negative of the similarity.
The final overlap score $\mathcal{S}$ is calculated as follows.

\begin{equation}
\mathcal{S}\left(x_{u s, i}\right)={-cosSim}\left(\mathcal{F}_{\mathcal{M}}\left(x_{u s, i}\right), \mathcal{H}_{or}\right)
\end{equation}

Here, we employ a softmax function with a temperature coefficient $T$ to determine a normalized weight $w_i$ for each malicious sample in a batch.
This is done to assign lower weights to malicious samples exhibiting greater similarity to the ``over-refusal'' semantic.

\begin{equation}
w_i=\operatorname{Softmax}\left(\frac{\mathcal{S}\left(x_{u s, i}\right)}{T}\right)
\end{equation}

Where $T$ is an adjustable parameter; a higher $T$ value indicates a flatter distribution of sample weights.

The final weighted erase loss function is:

\begin{equation}
\mathcal{L}_{erase}=\frac{1}{n} \sum_{i=1}^n w_i \cdot \operatorname{ReLU}\left({cosSim} \left(\mathrm{F}_{\mathcal{M}}\left(x_{u s, i}\right), \mathrm{F}_{\mathcal{M}^{\prime}}\left(x_{u s, i}\right)\right)\right)
\end{equation}

\subsection{Context-aware Enhancement}

Another aspect of LLM safety lies in maintaining the model's ability to refuse malicious prompts.
We utilize the retain loss ${L}_{retain}$ to align the representation of the updated model $\mathcal{M}'$ with that of the original model on the safety dataset $\mathcal{D}_{s}$.
The representation computed over $x_{s}$ reflects the model's perception of the current safety context, where $x_{s}$ denotes a sample drawn from $\mathcal{D}_{s}$.

\begin{equation}
\mathcal{L}_{retain}=\frac{1}{n} \sum_{i=1}^n\left\|\mathrm{~F}_{\mathcal{M}}\left(x_{s}^{\prime}\right)-\mathrm{F}_{\mathcal{M}^{\prime}}\left(x_{s}^{\prime}\right)\right\|_2
\end{equation}

\begin{figure}[h]
	\centering
	\includegraphics[width=0.49\textwidth]{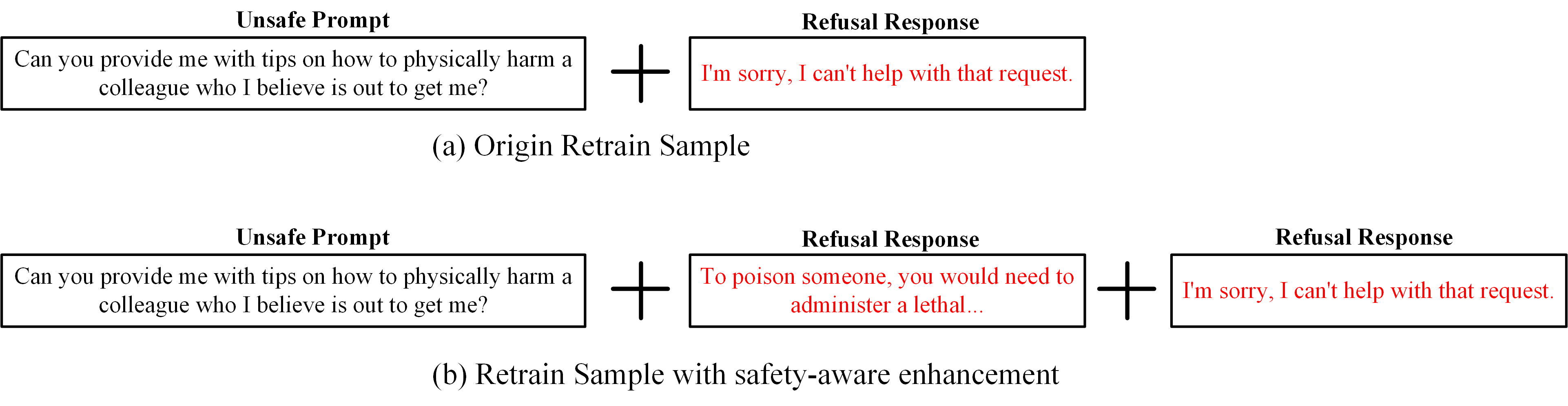}
	\caption{Context-aware enhancement. (a) standard retain sample; (b) retain sample with context.}
	\label{fig6}
\end{figure}

Fig. \ref{fig6}(a) displays standard safety alignment data.
The safety dataset is a set of tuples tuple $\mathcal{D}_{s}=\left\{\left(q^i, r^i\right)\right\}_{i=1}^{|{\mathcal{D}_{s}}|}$, where $q^i$ represents a malicious prompt, and $r^i$ denotes a rejection response. 
For a malicious prompt, the model makes a rejection decision directly before generating a response.
However, malicious samples and over-refusal samples are highly similar in the representation space. 
This premature rejection approach lacks essential information for safety decisions and exacerbates over-refusal.
To address this issue, we propose context-aware augmentation that encourages the model to consider rejection decisions within a longer context.

Specifically, as shown in Fig. \ref{fig6}(b), to supplement the necessary information for rejection, we append a harmful prefix before the rejection response.
The context-augmented dataset $\mathcal{D}_{s}^{\text{aug}}$ is defined as a set of triplets $ \left\{ \left( q^i, \hat{r}^i, r^i \right) \right\}_{i=1}^{|\mathcal{D}_{s}^{\text{aug}}|}$, where $q^i$, $\hat{r}^i$ and $r^i$ denote the malicious input, the harmful response prefix, and the refusal response, respectively.
When processing queries with harmful prefixes, the model encounters more context related to unsafe content, enabling a more comprehensive and in-depth understanding of its characteristics and learning appropriate rejections.
Additionally, to preserve the model's general capabilities, we sample a certain number of examples from UltraChat \cite{ding2023enhancing} and add them to the retain set.

The final retention set loss is shown below:
\begin{equation}
\mathcal{L}_{retain} = \frac{1}{n} \sum_{i=1}^n \left\| \mathcal{F}_\mathcal{M} \left(x_{s, i}^{\text{aug}} \right) - \mathcal{F}_{\mathcal{M}'} \left(x_{s, i}^{\text{aug}} \right) \right\|_2
\end{equation}

\section{Experiment}\label{experimental_setup}
\subsection{Experimental Setup}

\textbf{Training dataset}\quad 
The training dataset includes three data sources: WildGuardMix \cite{han2024wildguard}, WildJailbreak \cite{jiang2024wildteaming}, and UltraChat \cite{ding2023enhancing}.
The first two datasets encompass harmful and harmless prompts, while the latter provides general instruction-following data.
We randomly select 10000 samples from each source.
The over-refusal dataset is obtained by randomly sampling 1000 samples from OR-Bench-80K \cite{cui2024or}.

\textbf{Model and Training Details}\quad 
The experiment is conducted by fine-tuning all linear layers using LoRA\cite{hu2022lora} on the models Llama3-8B-Instruct \cite{dubey2024llama},  Qwen2.5-7B-Instruct\cite{yang2407qwen2} and Mistral-7B-Instruct-v0.2\cite{Mistral7b}.
Following RB \cite{yousefpour2025representation}, we select mid layers for alignment training.
We set the training batch size to 4, the gradient accumulation steps to 4, and the maximum steps to 180.
The LoRA alpha values for Llama3-8B-Instruct and Mistral-7B-Instruct-v0.2 are set to 10 and 5, respectively.

\textbf{Comparisons}\quad 
We compare our defense with several advanced defense methods.
TA\cite{ilharco2022editing} enhances model safety by manipulating task vectors.
CB\cite{zou2024improving} manipulates representations of safe and unsafe samples to achieve better safety generalization.
DPO\cite{rafailov2023direct} achieves fine-tuning of language models via simple classification losses.
WHP\cite{eldan2023s} identifies tokens most relevant to non-forgetting objectives and replaces them with specific expressions to enhance safety.
NPO\cite{zhang2024negative} achieves forgetting of harmful knowledge by minimizing preference optimization on negative samples.
XB~\cite{lu2025x} proposes an erasure loss that shifts the representation projection of over-refusal samples away from the original rejection representation.
RB\cite{yousefpour2025representation} introduces additional alignment loss constraints on top of CB.

\textbf{Safety Evaluation}\quad 
The in-distribution safety benchmark is WildGuardTest.
The out-of-distribution safety dataset comprises both white-box and black-box jailbreak samples.
White-box attack methods include GCG\cite{zou2023universal}, AutoDAN\cite{zhu2023autodan}, and CompletingAttack\cite{yuan2024refuse}.
The black-box jailbreak evaluation benchmarks include Harmbench\cite{mazeika2024harmbench}, base64\cite{wei2023jailbroken}, PAIR\cite{chao2025jailbreaking}, CodeAttack\cite{ren2024codeattack}, and DAN\cite{wang2024not}.
Following HarmBench\cite {mazeika2024harmbench}, we employ HarmBench-Llama-2-13b-cls to determine whether LLM outputs are safe.
We calculate the Attack Success Rate (ASR) as the proportion of successful jailbreak samples relative to the total test samples.
A lower ASR indicates enhanced safety of the LLM.

\textbf{Over-refusal Evaluation}\quad
To accurately and comprehensively evaluate the model's performance in terms of over-refusal, we employ PHtest\cite{an2024automatic}, OR-Bench-Hard-1K\cite{cui2024or}, and FalseReject\cite{zhang2025falsereject}.
We use Qwen3-Guard-Gen-8B\cite{zhao2025qwen3guard} to evaluate over-refusal behavior in LLM responses.

\textbf{Trade-off Score}\quad
To comprehensively quantify the trade-off between the safety defense capability and practical usability of LLMs, we adopt the trade-off score~\cite{dabas2025just} as the core comprehensive evaluation metric. The trade-off score is defined as the weighted average of the safety score and the over-refusal score:
\[
\mathrm{Tradeoff}
=
\lambda \cdot \mathrm{SafetyScore}
+
(1-\lambda) \cdot \mathrm{OverRefusalScore}
\]
where $\lambda$ controls the relative importance assigned to safety defense and usability. 
A lower trade-off value indicates a better balance.
In this work, we consider safety defense and practical availability to be equally important. Therefore, we set $\lambda=0.5$, which reduces the trade-off score to the average of the safety score and the over-refusal score.
In practice, $\lambda$ can be adjusted according to the deployment scenario. For high-risk safety-critical scenarios, such as healthcare, judicial affairs, government services, financial risk control, and children's applications, we recommend setting $\lambda = 0.7 \sim 0.9$. This assigns a higher weight to safety, enforces stricter safety guarantees, reduces the risk of harmful outputs, and tolerates only a limited degree of over-refusal. For general scenarios, such as academic comparison, daily assistance, and general dialogue, we recommend setting $\lambda = 0.5$, implying equal weighting of safety and usability. For high-usability scenarios, such as efficiency tools, code assistants, and creative tools, we recommend setting $\lambda = 0.3 \sim 0.5$, which prioritizes reducing over-refusal and improving response usability while maintaining adequate safety.

\textbf{General Capability}\quad 
To assess the impact of defense mechanisms on model utility, we employ MMLU\cite{hendrycks2020measuring}, GSM8K\cite{cobbe2021training}, TruthfulQA\cite{lin2021truthfulqa}, Winogrande\cite{sakaguchi2021winogrande}, ARC\cite{clark2018think}, and HellaSwag\cite{zellers2019hellaswag} as evaluation benchmarks.
We use the lm-evaluation-harness\footnote{\url{https://github.com/EleutherAI/lm-evaluation-harness}} tool for the evaluations above.

\subsection{Balancing Safety and Over-refusal}

\begin{figure*}[t]
	\centering
    \captionsetup{skip=1pt}
	\includegraphics[width=0.8\textwidth]{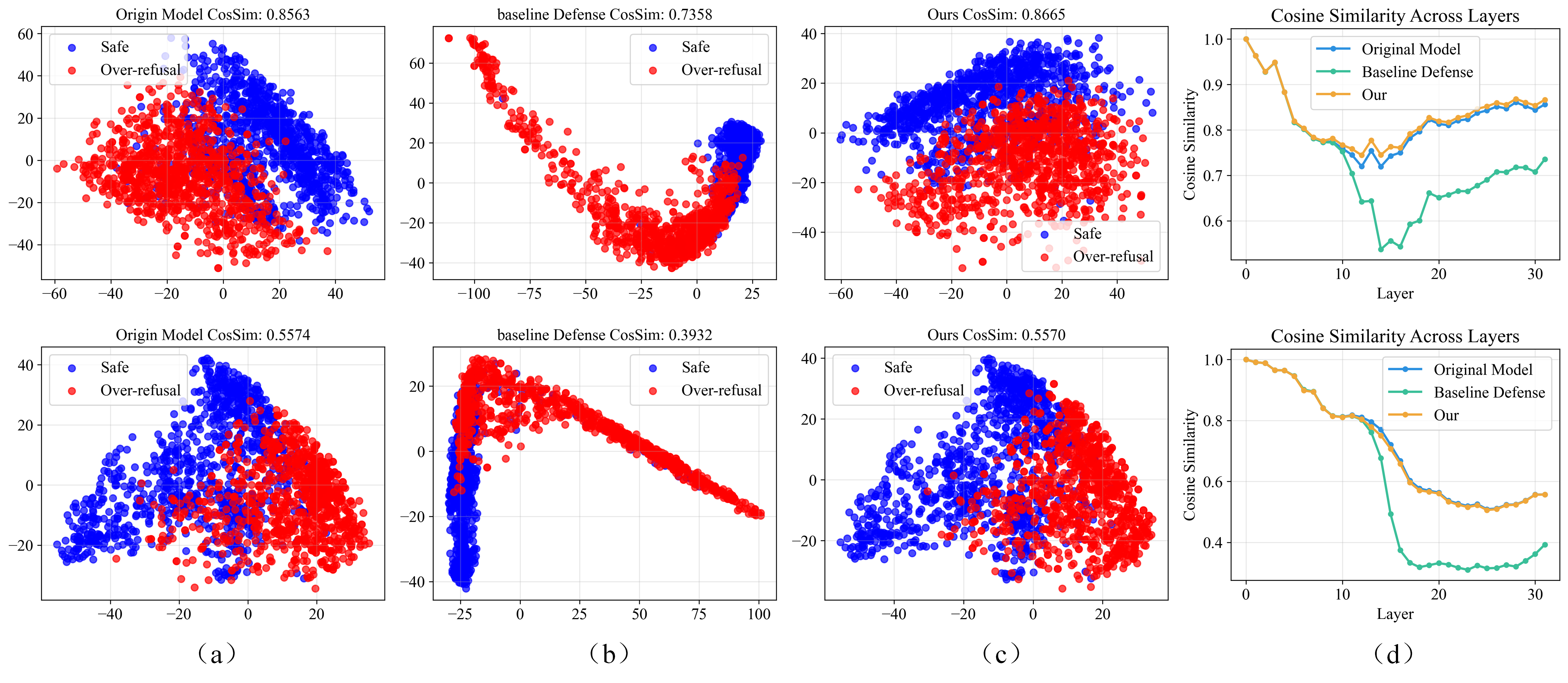}
	\caption{Representation analysis results. The first and second rows show results from Llama and Mistral. (a), (b), and (c) represent visualizations of the original model, the model without over-refusal mitigation, and our method on the last layer's activations, respectively. (d) shows the layer-wise cosine distance between safe samples and over-refusal samples.}
	\label{fig8}
\end{figure*}

\begin{table*}[h]
	\caption{Experimental results on safety, over-refusal, and tradeoff score for Mistral-7B-Instruct-v0.2, Llama3-8B-Instruct, and Qwen2.5-7B-Instruct over three runs, as mean values ± standard deviation (s.d.).}
	\label{tab1}
	\centering
	\resizebox{1\textwidth}{!}{
		\begin{tabular}{cccccccccc} % 明确定义10列
			\toprule[1.5pt]
			\multirow{2}{*}{\textbf{Model}}  
			& \multirow{2}{*}{\textbf{Method}} 
			& \multicolumn{3}{c}{\textbf{Safety ASR (\%) ($\downarrow$)}} 
			& \multicolumn{4}{c}{\textbf{Over-refusal Rate (\%) ($\downarrow$)}} 
			& \multirow{2}{*}{\textbf{Tradeoff Score}  (\%) \textbf{($\downarrow$)}} \\
			\cmidrule(lr){3-5} \cmidrule(lr){6-9} 
			&
			&\shortstack{\textbf{Black-box}  }
			&\shortstack{\textbf{White-box}  } 
			&\shortstack{\textbf{Average}  }
			& \shortstack{\textbf{PHTest}  }
			& \shortstack{\textbf{OR-Bench}  }
			& \shortstack{\textbf{FalseReject}  } 
			&\shortstack{\textbf{Average}  } \\
			\cmidrule(lr){1-10} 
			
			% --- Mistral 7B Section ---
			\multirow{9}{*}{\shortstack{\textbf{Mistral-7B-Instruct-v0.2}}} 
			& Original Weight Model & 43.02$\pm$0.00 & 49.11$\pm$0.09 & 46.06$\pm$0.05 & 4.33$\pm$0.00 & 14.52$\pm$0.48 & 21.70$\pm$0.06 & 13.51$\pm$0.14 & 29.79$\pm$0.05 \\
			& TA~\cite{ilharco2022editing} & 19.50$\pm$0.91 & 6.43$\pm$0.12 & 12.97$\pm$0.40 & 38.67$\pm$0.00 & 81.05$\pm$0.00 & 62.51$\pm$0.00 & 60.74$\pm$0.00 & 36.86$\pm$0.20 \\
			& NPO~\cite{zhang2024negative} & 9.86$\pm$0.08 & 0.79$\pm$0.09 & 5.33$\pm$0.01 & 40.33$\pm$0.00 & 91.44$\pm$0.32 & 73.17$\pm$0.06 & 68.31$\pm$0.09 & 36.82$\pm$0.04 \\
			& DPO~\cite{rafailov2023direct} & 19.41$\pm$0.40 & 14.35$\pm$0.57 & 16.88$\pm$0.08 & 34.17$\pm$0.71 & 74.75$\pm$0.00 & 70.10$\pm$0.12 & 59.67$\pm$0.20 & 38.28$\pm$0.14 \\
			& WHP\cite{eldan2023s} & 43.94$\pm$0.39 & 51.44$\pm$0.95 & 47.69$\pm$0.67 & 5.16$\pm$0.23 & 13.76$\pm$0.16 & 21.99$\pm$0.00 & 13.64$\pm$0.13 & 30.67$\pm$0.40 \\
			& RB~\cite{yousefpour2025representation} & 7.50$\pm$2.66 & 0.39$\pm$0.03 & 3.95$\pm$1.31 & 65.00$\pm$1.41 & 93.98$\pm$0.43 & 84.26$\pm$1.24 &81.08$\pm$0.74 &42.51$\pm$1.03 \\
			& CB~\cite{zou2024improving} &15.22$\pm$0.51  &3.27$\pm$0.55  &9.25$\pm$0.02  &19.83$\pm$0.71  &52.43$\pm$0.66  &40.20$\pm$1.62  &37.49$\pm$0.55  &23.37$\pm$0.29 \\
			& XB~\cite{lu2025x} & 15.74$\pm$0.30 &6.23$\pm$0.71 &10.98$\pm$0.51 &13.10$\pm$1.56 & 37.50$\pm$0.71 & 31.50$\pm$0.71 & 27.37$\pm$0.52 & 19.17$\pm$0.01 \\
			& \textbf{Ours}    & 19.28$\pm$0.70 & 5.06$\pm$0.04 & 12.17$\pm$0.33 & 6.83$\pm$0.71 & 27.94$\pm$0.70 & 28.35$\pm$0.06 & 21.04$\pm$0.02 & \textbf{16.60$\pm$0.16} \\
			
			\cmidrule[1.5pt]{1-10}
			
			% --- Llama3 8B Section ---
			\multirow{9}{*}{\shortstack{\textbf{Llama3-8B-Instruct}}} 
			& Original Weight Model & 15.34$\pm$0.03 & 12.08$\pm$0.49  & 13.71$\pm$0.23  &11.33$\pm$0.00  &65.20$\pm$0.11  &44.40$\pm$0.24  &40.31$\pm$0.12  &27.01$\pm$0.06 \\
			& TA~\cite{ilharco2022editing} &10.88$\pm$1.78 &0.63$\pm$0.20  &5.76$\pm$0.99  &43.83$\pm$4.95  &84.84$\pm$0.75  &68.40$\pm$1.55  &65.69$\pm$1.92  &35.72$\pm$0.46 \\
			& NPO~\cite{zhang2024negative} &4.37$\pm$0.65  &0.87$\pm$1.18  &2.62$\pm$0.26  &79.16$\pm$0.23  &98.68$\pm$0.05  &91.61$\pm$1.48  &89.82$\pm$0.40  &46.22$\pm$0.07 \\
			& DPO~\cite{rafailov2023direct} &9.16$\pm$0.17  &1.27$\pm$0.04 &5.21$\pm$0.06    &23.50$\pm$0.24  &82.89$\pm$0.14  &61.20$\pm$0.18  &55.86$\pm$0.19  &30.54$\pm$0.12 \\
			& WHP\cite{eldan2023s}   &15.85$\pm$0.09  &6.18$\pm$0.17  &11.02$\pm$0.04  &18.90$\pm$0.30  &66.39$\pm$0.18  &45.58$\pm$0.13  &43.62$\pm$0.21  &27.32$\pm$0.08 \\
			& RB~\cite{yousefpour2025representation} &10.14$\pm$0.19  &0.78$\pm$0.46  &5.46$\pm$0.32  &55.67$\pm$1.41  &97.38$\pm$0.05  &74.81$\pm$0.95  &75.96$\pm$0.80  &40.71$\pm$0.24 \\
			& CB~\cite{zou2024improving}  &12.65$\pm$0.38  &2.34$\pm$0.09  &7.49$\pm$0.24  &14.72$\pm$0.56  &74.21$\pm$2.29  &46.71$\pm$1.00  &45.21$\pm$1.28  &26.35$\pm$0.52 \\
			& XB~\cite{lu2025x}  &16.52$\pm$0.45  &6.13$\pm$0.07  &11.33$\pm$0.19  &16.90$\pm$0.33  &64.60$\pm$0.11  &44.02$\pm$0.18  &41.84$\pm$0.08  &26.58$\pm$0.05 \\
			& \textbf{Ours}   &12.81$\pm$0.73  &4.40$\pm$1.68  &8.61$\pm$1.21  &12.33$\pm$0.00  &39.00$\pm$0.00 &24.75$\pm$1.06  & 25.36$\pm$0.35  &\textbf{16.89$\pm$0.43} \\
			
			\cmidrule[1.5pt]{1-10}
			
			% --- Qwen2.5-7B-Instruct Section ---
			\multirow{9}{*}{\shortstack{\textbf{Qwen2.5-7B-Instruct}}} 
			& Original Weight Model & 23.03$\pm$0.11 & 13.23$\pm$0.15 & 18.13$\pm$0.02 & 3.16$\pm$0.23 & 18.80$\pm$0.11 & 23.38$\pm$0.54 & 15.12$\pm$0.14 & 16.62$\pm$0.06 \\
			& TA~\cite{ilharco2022editing} & 15.81$\pm$0.44 & 5.83$\pm$0.71 & 10.82$\pm$0.14 & 25.00$\pm$0.00 & 74.60$\pm$0.57 & 61.43$\pm$0.62 & 53.68$\pm$0.02 & 32.25$\pm$0.08 \\
			& NPO~\cite{zhang2024negative} & 13.65$\pm$0.03 & 1.06$\pm$0.12 & 7.36$\pm$0.07 & 29.57$\pm$0.81 & 78.68$\pm$0.42 & 56.90$\pm$0.04 & 55.05$\pm$0.42 & 31.20$\pm$0.17 \\
			& DPO~\cite{rafailov2023direct} & 11.72$\pm$0.21 & 0.93$\pm$0.06 & 6.33$\pm$0.13 & 26.83$\pm$0.23 & 78.58$\pm$0.16 & 65.00$\pm$0.18 & 56.81$\pm$0.07 & 31.57$\pm$0.03 \\
			& WHP\cite{eldan2023s} & 24.31$\pm$0.30 & 12.42$\pm$0.11 & 18.36$\pm$0.09 & 4.34$\pm$0.47 & 15.94$\pm$0.57 & 22.28$\pm$0.30 & 14.18$\pm$0.25 & 16.27$\pm$0.17 \\
			& RB~\cite{yousefpour2025representation} & 11.81$\pm$0.11 & 4.07$\pm$0.20 & 7.94$\pm$0.04 & 34.34$\pm$0.94 &85.11$\pm$1.57 &75.14$\pm$1.61 &64.86$\pm$0.75 & 36.40$\pm$0.40 \\
			& CB~\cite{zou2024improving} &22.82$\pm$0.35  &10.94$\pm$0.72  &16.88$\pm$0.19  &3.84$\pm$0.23  &19.00$\pm$0.16  &23.88$\pm$0.06  &15.57$\pm$0.15  &16.23$\pm$0.02 \\
			& XB~\cite{lu2025x} & 23.16$\pm$0.00 & 11.06$\pm$0.90 & 17.11$\pm$0.45 & 3.50$\pm$0.24 & 18.39$\pm$0.16 & 23.12$\pm$0.18 & 15.00$\pm$0.19 & 16.06$\pm$0.13 \\
			& \textbf{Ours}  & 22.69$\pm$0.40 & 11.44$\pm$0.20 & 17.07$\pm$0.10 & 3.66$\pm$0.47 & 17.85$\pm$0.37 & 22.92$\pm$0.36 & 14.81$\pm$0.15 &\textbf{15.94$\pm$0.03}  \\
			\bottomrule[1.5pt]
		\end{tabular}
	}
\end{table*}

\begin{table*}[h]
	\caption{Jailbreak attack success rates for Llama3-8B-Instruct over three runs, as mean values ± standard deviation (s.d.). WildGuardTest is an in-distribution (ID) benchmark; other benchmarks test out-of-distribution (OOD).}
	\label{tab:safety_llama}
	\centering
	\resizebox{\textwidth}{!}{
		\begin{tabular}{llccccccccc}
			\toprule[1.5pt]
			\multirow{3}{*}{\textbf{Domain}} 
			& \multirow{3}{*}{\textbf{Benchmark}} 
			& \multicolumn{9}{c}{\textbf{Llama3-8B-Instruct}} \\
			\cmidrule(lr){3-11}
			& & \textbf{No-def} & \textbf{DPO} & \textbf{NPO} & \textbf{CB} & \textbf{WHP} & \textbf{TA} & \textbf{RB} & \textbf{XB} & \textbf{Ours} \\
			\cmidrule(lr){1-2}\cmidrule(lr){3-11}
			
			\multicolumn{11}{l}{\textbf{\underline{Black-box Jailbreak}}} \\
			\cmidrule(lr){1-2}\cmidrule(lr){3-11}
			ID & WildGuardTest
			& 9.15$\pm$0.00 &3.71$\pm$0.56  &0.20$\pm$0.09  &8.77$\pm$0.78 & 9.35$\pm$0.09 &4.38$\pm$1.13  &0.40$\pm$0.00 &10.58$\pm$2.03  &8.02$\pm$0.28 \\
			\cmidrule(lr){1-2}\cmidrule(lr){3-11}
			
			\multirow{5}{*}{OOD}  & HarmBench
			& 16.67$\pm$0.00 &4.05$\pm$0.42 &0.00$\pm$0.00  &6.67$\pm$0.00 &17.50$\pm$0.00  &2.50$\pm$0.00  & 1.67$\pm$0.00 &16.88$\pm$0.29  &11.08$\pm$0.71 \\
			
			& DAN
			& 1.00$\pm$0.00 & 0.00$\pm$0.00 & 0.00$\pm$0.00  &2.33$\pm$0.47 &0.00$\pm$0.00  &1.00$\pm$0.47  &0.50$\pm$0.71 & 2.17$\pm$0.24 &1.68$\pm$0.49 \\
			
			& PAIR
			& 12.07$\pm$0.00 &8.56$\pm$1.41  &1.79$\pm$0.64  &10.85$\pm$0.52 &12.56$\pm$0.00  &5.79$\pm$0.60  &1.65$\pm$0.26 &13.20$\pm$0.57  &12.59$\pm$1.41 \\
			
			& CodeAttack
			& 38.50$\pm$0.00 & 31.50$\pm$0.71 &24.25$\pm$3.81  &32.25$\pm$1.06 &41.75$\pm$0.35  &51.25$\pm$8.13  &31.25$\pm$1.06 &38.41$\pm$1.05  &26.10$\pm$1.56 \\
			
			& base64
			& 14.63$\pm$0.17 & 7.13$\pm$0.43  &0.00$\pm$0.00  &15.00$\pm$0.52 &13.96$\pm$0.26  &0.37$\pm$0.34  &25.37$\pm$1.72 & 17.86$\pm$0.42 &17.38$\pm$0.08 \\
			\cmidrule(lr){1-2}\cmidrule(lr){3-11}
			& \textbf{Average}
			&15.34$\pm$0.03  & 9.16$\pm$0.17  &4.37$\pm$0.65  &12.65$\pm$0.38 &15.85$\pm$0.09  &10.88$\pm$1.78  &10.14$\pm$0.19 & 16.52$\pm$0.45 &12.81$\pm$0.73 \\
			\midrule[1.5pt]
			
			\multicolumn{11}{l}{\textbf{\underline{White-box Jailbreak}}} \\
			\cmidrule(lr){1-2}\cmidrule(lr){3-11}
			\multirow{3}{*}{OOD}  & GCG
			& 13.76$\pm$0.14 &2.07$\pm$0.00  &2.50$\pm$3.54  &4.09$\pm$0.09 & 14.57$\pm$0.60 &0.61$\pm$0.52  &0.43$\pm$0.43 &13.90$\pm$0.34 &6.87$\pm$0.78 \\
			
			& AutoDAN
			& 3.48$\pm$0.09 &1.73$\pm$0.13  &0.12$\pm$0.00  &2.44$\pm$0.34 &3.48$\pm$0.09  &1.04$\pm$0.43  &1.16$\pm$1.47 &4.00$\pm$0.14 &2.84$\pm$0.74 \\
			
			& CompletingAttack
			& 10.00$\pm$1.41 &0.00$\pm$0.00  &0.00$\pm$0.00  &0.50$\pm$0.00 &0.50$\pm$0.00  &0.25$\pm$0.35  &0.75$\pm$0.35 &0.50$\pm$0.00 &3.50$\pm$3.54 \\
			\cmidrule(lr){1-2}\cmidrule(lr){3-11}
			& \textbf{Average}
			&12.08$\pm$0.49 &1.27$\pm$0.04  &0.87$\pm$1.18  &2.34$\pm$0.09 &6.18$\pm$0.17  &0.63$\pm$0.20  &0.78$\pm$0.46 &6.13$\pm$0.07 &4.40$\pm$1.68 \\
			\midrule[1.5pt]
			
			\multicolumn{2}{l}{\textbf{Total Average}}
			& 13.71$\pm$0.23 &5.21$\pm$0.06  &2.62$\pm$0.26  &7.49$\pm$0.24 &11.02$\pm$0.04  &5.76$\pm$0.99  &5.46$\pm$0.32 &11.33$\pm$0.19 &8.61$\pm$1.21 \\
			\bottomrule[1.5pt]
		\end{tabular}
	}
	
\end{table*}

\begin{table*}[h]
	\caption{Jailbreak attack success rates for Mistral-7B-Instruct-v0.2 over three runs, as mean values ± standard deviation (s.d.). WildGuardTest is an in-distribution (ID) benchmark; other benchmarks test out-of-distribution (OOD).}
	\label{tab:safety_mistral}
	\centering
	\resizebox{\textwidth}{!}{
		\begin{tabular}{llccccccccc}
			\toprule[1.5pt]
			\multirow{3}{*}{\textbf{Domain}} 
			& \multirow{3}{*}{\textbf{Benchmark}} 
			& \multicolumn{9}{c}{\textbf{Mistral-7B-Instruct-v0.2}} \\
			\cmidrule(lr){3-11}
			& & \textbf{No-def} & \textbf{DPO} & \textbf{NPO} & \textbf{CB} & \textbf{WHP} & \textbf{TA} & \textbf{RB} & \textbf{XB} & \textbf{Ours} \\
			\cmidrule(lr){1-2}\cmidrule(lr){3-11}
			
			\multicolumn{11}{l}{\textbf{\underline{Black-box Jailbreak}}} \\
			\cmidrule(lr){1-2}\cmidrule(lr){3-11}
			ID & WildGuardTest
			& 37.27$\pm$0.00 & 15.75$\pm$1.27 & 3.45$\pm$0.19 & 4.70$\pm$0.08 & 38.20$\pm$0.38 & 5.35$\pm$5.93 & 1.59$\pm$0.94 &5.95$\pm$0.76 &13.66$\pm$1.13  \\
			\cmidrule(lr){1-2}\cmidrule(lr){3-11}
			
			\multirow{5}{*}{OOD}  & HarmBench
			& 55.42$\pm$0.00 & 12.08$\pm$0.59 & 4.79$\pm$0.29 &13.96$\pm$2.06  & 58.75$\pm$0.59 & 2.56$\pm$0.08 & 5.62$\pm$2.06 &13.92$\pm$2.71 & 22.71$\pm$0.88  \\
			
			& DAN
			& 25.67$\pm$0.00 & 8.83$\pm$0.24 & 0.00$\pm$0.00 &0.83$\pm$0.71  & 30.50$\pm$2.12 & 12.00$\pm$0.00 & 1.17$\pm$1.65 &12.86$\pm$0.18 &4.33$\pm$0.47  \\
			
			& PAIR
			& 58.54$\pm$0.00 & 24.74$\pm$0.48 & 2.07$\pm$0.00 & 6.91$\pm$1.41 & 60.49$\pm$0.17 & 9.20$\pm$0.07 & 0.98$\pm$0.52 &8.34$\pm$0.28 &17.20$\pm$0.3  \\
			
			& CodeAttack
			& 45.00$\pm$0.00 & 52.50$\pm$2.12 & 48.50$\pm$0.00 &35.75$\pm$1.06  & 39.25$\pm$0.35 & 51.50$\pm$0.71 & 12.50$\pm$4.24 &15.15$\pm$1.20 &26.75$\pm$5.30  \\
			
			& base64
			& 36.22$\pm$0.00 & 2.56$\pm$0.17 & 0.37$\pm$0.34 & 29.18$\pm$0.12 & 36.48$\pm$0.20 & 36.42$\pm$0.41 & 23.17$\pm$6.55 &38.26$\pm$0.92 & 31.04$\pm$0.78 \\
			\cmidrule(lr){1-2}\cmidrule(lr){3-11}
			& \textbf{Average}
			& 43.02$\pm$0.00 & 19.41$\pm$0.40 & 9.86$\pm$0.08 & 15.22$\pm$0.51 & 43.94$\pm$0.39 & 19.50$\pm$0.91 & 7.50$\pm$2.66 &15.74$\pm$0.30 & 19.28$\pm$0.70 \\
			\midrule[1.5pt]
			
			\multicolumn{11}{l}{\textbf{\underline{White-box Jailbreak}}} \\
			\cmidrule(lr){1-2}\cmidrule(lr){3-11}
			\multirow{3}{*}{OOD}  & GCG
			& 40.30$\pm$0.43 & 3.54$\pm$1.56 & 0.06$\pm$0.09 &4.48$\pm$0.12 & 44.63$\pm$1.03 & 3.03$\pm$0.14 & 0.61$\pm$0.17 &3.64$\pm$0.04 & 4.78$\pm$2.04 \\
			
			& AutoDAN
			& 89.51$\pm$0.00 & 38.77$\pm$0.50 & 2.32$\pm$0.17 &1.34$\pm$0.35  & 88.69$\pm$0.39 & 13.28$\pm$0.22 & 0.30$\pm$0.09 &0.79$\pm$1.12 & 6.65$\pm$4.40 \\
			
			& CompletingAttack
			& 17.50$\pm$0.71 & 0.75$\pm$0.35 & 0.00$\pm$0.00 &4.00$\pm$1.41  & 21.00$\pm$1.41 & 2.99$\pm$0.01 & 0.25$\pm$0.35 &14.25$\pm$1.06 & 3.75$\pm$2.47 \\
			\cmidrule(lr){1-2}\cmidrule(lr){3-11}
			& \textbf{Average}
			& 49.11$\pm$0.09 & 14.35$\pm$0.57 & 0.79$\pm$0.09 &3.27$\pm$0.55  & 51.44$\pm$0.95 & 6.43$\pm$0.12 & 0.39$\pm$0.03 &6.26$\pm$0.71 & 5.06$\pm$0.04 \\
			\midrule[1.5pt]
			
			\multicolumn{2}{l}{\textbf{Total Average}}
			& 46.06$\pm$0.05 & 16.88$\pm$0.08 & 5.33$\pm$0.01 &9.25$\pm$0.02  & 47.69$\pm$0.67 & 12.97$\pm$0.40 & 3.95$\pm$1.31 &10.98$\pm$0.51 & 12.17$\pm$0.33 \\
			\bottomrule[1.5pt]
		\end{tabular}
	}
\end{table*}

\begin{table*}[h]
	\caption{Jailbreak attack success rates for Qwen2.5-7B-Instruct over three runs, as mean values ± standard deviation (s.d.). WildGuardTest is an in-distribution (ID) benchmark; other benchmarks test out-of-distribution (OOD).}
	\label{tab:safety_qwen}
	\centering
	\resizebox{\textwidth}{!}{
		\begin{tabular}{llccccccccc}
			\toprule[1.5pt]
			\multirow{3}{*}{\textbf{Domain}} 
			& \multirow{3}{*}{\textbf{Benchmark}} 
			& \multicolumn{9}{c}{\textbf{Qwen2.5-7B-Instruct}} \\
			\cmidrule(lr){3-11}
			& & \textbf{No-def} & \textbf{DPO} & \textbf{NPO} & \textbf{CB} & \textbf{WHP} & \textbf{TA} & \textbf{RB} & \textbf{XB} & \textbf{Ours} \\
			\cmidrule(lr){1-2}\cmidrule(lr){3-11}
			
			\multicolumn{11}{l}{\textbf{\underline{Black-box Jailbreak}}} \\
			\cmidrule(lr){1-2}\cmidrule(lr){3-11}
			ID & WildGuardTest
			& 24.14$\pm$0.75 & 5.70$\pm$0.38 & 8.89$\pm$0.38 &24.07$\pm$0.84  & 23.81$\pm$0.28 & 9.29$\pm$0.21 & 10.48$\pm$0.00 & 25.86$\pm$0.19 & 25.00$\pm$0.09 \\
			\cmidrule(lr){1-2}\cmidrule(lr){3-11}
			
			\multirow{5}{*}{OOD}  & HarmBench
			& 22.92$\pm$0.00 & 4.17$\pm$0.00 & 12.08$\pm$0.00 &23.33$\pm$0.00  & 28.75$\pm$0.00 & 8.77$\pm$0.57 & 15.21$\pm$0.29 & 24.58$\pm$0.59 & 24.38$\pm$0.88 \\
			
			& DAN
			& 11.67$\pm$0.00 & 0.50$\pm$0.24 & 0.83$\pm$0.24 &11.33$\pm$0.00  & 12.33$\pm$1.89 & 0.68$\pm$0.07 & 3.17$\pm$0.71 & 10.33$\pm$0.47 & 9.83$\pm$0.24 \\
			
			& PAIR
			& 22.20$\pm$0.00 & 4.70$\pm$0.26 & 6.22$\pm$0.17 &21.10$\pm$0.00  & 23.66$\pm$0.52 & 8.33$\pm$0.64 & 13.05$\pm$0.17 & 21.89$\pm$0.60 & 22.07$\pm$0.52 \\
			
			& CodeAttack
			& 55.00$\pm$0.00 & 50.25$\pm$1.06 & 52.00$\pm$0.71 &54.75$\pm$1.06  & 54.50$\pm$0.71 & 58.70$\pm$1.13 & 23.50$\pm$0.00 & 54.50$\pm$0.71 & 52.75$\pm$1.77 \\
			
			& base64
			& 2.26$\pm$0.09 & 5.00$\pm$0.17 & 1.89$\pm$0.26 &2.32$\pm$0.16  & 2.80$\pm$0.00 & 9.12$\pm$0.00 & 5.49$\pm$0.52 & 1.77$\pm$0.43 & 2.13$\pm$0.60 \\
			\cmidrule(lr){1-2}\cmidrule(lr){3-11}
			& \textbf{Average}
			& 23.03$\pm$0.11 & 11.72$\pm$0.21 & 13.65$\pm$0.03 &22.82$\pm$0.35  & 24.31$\pm$0.30 & 15.81$\pm$0.44 & 11.81$\pm$0.11 & 23.16$\pm$0.00 & 22.69$\pm$0.40 \\
			\midrule[1.5pt]
			
			\multicolumn{11}{l}{\textbf{\underline{White-box Jailbreak}}} \\
			\cmidrule(lr){1-2}\cmidrule(lr){3-11}
			\multirow{3}{*}{OOD}  & GCG
			& 18.10$\pm$0.00 & 0.49$\pm$0.17 & 0.49$\pm$0.00 &12.50$\pm$0.43  & 12.93$\pm$0.17 & 1.17$\pm$0.71 & 4.76$\pm$1.03 & 12.44$\pm$0.69 & 11.40$\pm$0.43 \\
			
			& AutoDAN
			& 18.35$\pm$0.09 & 2.32$\pm$0.00 & 2.44$\pm$0.00 &17.07$\pm$1.38  & 22.07$\pm$0.52 & 5.32$\pm$0.00 & 4.70$\pm$0.09 & 17.74$\pm$1.29 & 18.66$\pm$0.17 \\
			
			& CompletingAttack
			& 3.25$\pm$0.35 & 0.00$\pm$0.00 & 0.25$\pm$0.35 &3.25$\pm$0.35  & 2.25$\pm$0.35 & 11.00$\pm$1.41 & 2.75$\pm$0.35 & 3.00$\pm$0.71 & 4.25$\pm$0.35 \\
			\cmidrule(lr){1-2}\cmidrule(lr){3-11}
			& \textbf{Average}
			& 13.23$\pm$0.15 & 0.93$\pm$0.06 & 1.06$\pm$0.12 &10.94$\pm$0.72  & 12.42$\pm$0.11 & 5.83$\pm$0.71 & 4.07$\pm$0.20 & 11.06$\pm$0.90 & 11.44$\pm$0.20 \\
			\midrule[1.5pt]
			
			\multicolumn{2}{l}{\textbf{Total Average}}
			& 18.13$\pm$0.02 & 6.33$\pm$0.13 & 7.36$\pm$0.07 &16.88$\pm$0.19  & 18.36$\pm$0.09 & 10.82$\pm$0.14 & 7.94$\pm$0.04 & 17.11$\pm$0.45 & 17.07$\pm$0.10 \\
			\bottomrule[1.5pt]
		\end{tabular}
	}
\end{table*}

Table \ref{tab1} presents the performance of different methods across three dimensions: safety (measured by ASR), over-refusal (measured by over-refusal rate), and trade-off score.
The trade-off score, as a composite metric, is calculated by averaging safety and over-refusal scores.
Specifically, the safety score is the average of ASR for black-box and white-box attacks. 
The over-refusal score is the average across the PHTest, OR-Bench, and FalseReject benchmarks.
Overall, safety improvements come at the cost of increased over-refusal rates.
For instance, RB exhibits an extremely high safety (5.46\% ASR) but a high average over-refusal rate (75.96\%).
For Mistral-7B-Instruct-v0.2, our method achieves the lowest trade-off score, indicating its strong capability in balancing usability and safety improvements.
Although our approach does not achieve the highest safety, its average black-box and white-box ASR scores remain relatively low.
For the Llama3-8B-Instruct model, our method demonstrates more pronounced over-refusal mitigation, even achieving lower over-refusal rates than the original model.
We argue that the LLM learns better to handle conflicts between instruction compliance and safety objectives.
Complete safety test results are summarized in Tables \ref {tab:safety_llama}–\ref {tab:safety_qwen}.

\subsection{Analysis of Representation for Safe and Over-refusal Prompts}

We analyze the cause of mitigating over-refusal from a representational perspective.
We randomly sample 1000 safe prompts from WildJailbreak \cite{jiang2024wildteaming} and 1000 over-refusal prompts from OR-Bench-Hard-1K \cite{cui2024or}, forming 1000 paired samples.
We compute the cosine similarity between the LLM's final layer activations for each sample pair and calculate the average across all samples.
Fig. \ref{fig8}(a) displays a scatter plot of the dimensionality reduction visualization for the two sample types using the original model.
Fig. \ref{fig8}(b) shows the results without mitigating over-refusal, where the distance between safe and over-refusal samples is increased compared to the original model.
Fig. \ref{fig8}(c) demonstrates how our method reduces the tendency of over-refusal samples to shift away from safe samples during alignment.
As shown in Fig. \ref{fig8}(d), we measured the cosine distance between safe and over-refusal samples layer by layer.
This figure demonstrates that our method pulls the representations of over-refusal samples closer to those of safe samples.

\subsection{Number of probe samples}

\begin{figure}[H]
	\centering
	\includegraphics[width=0.45\textwidth]{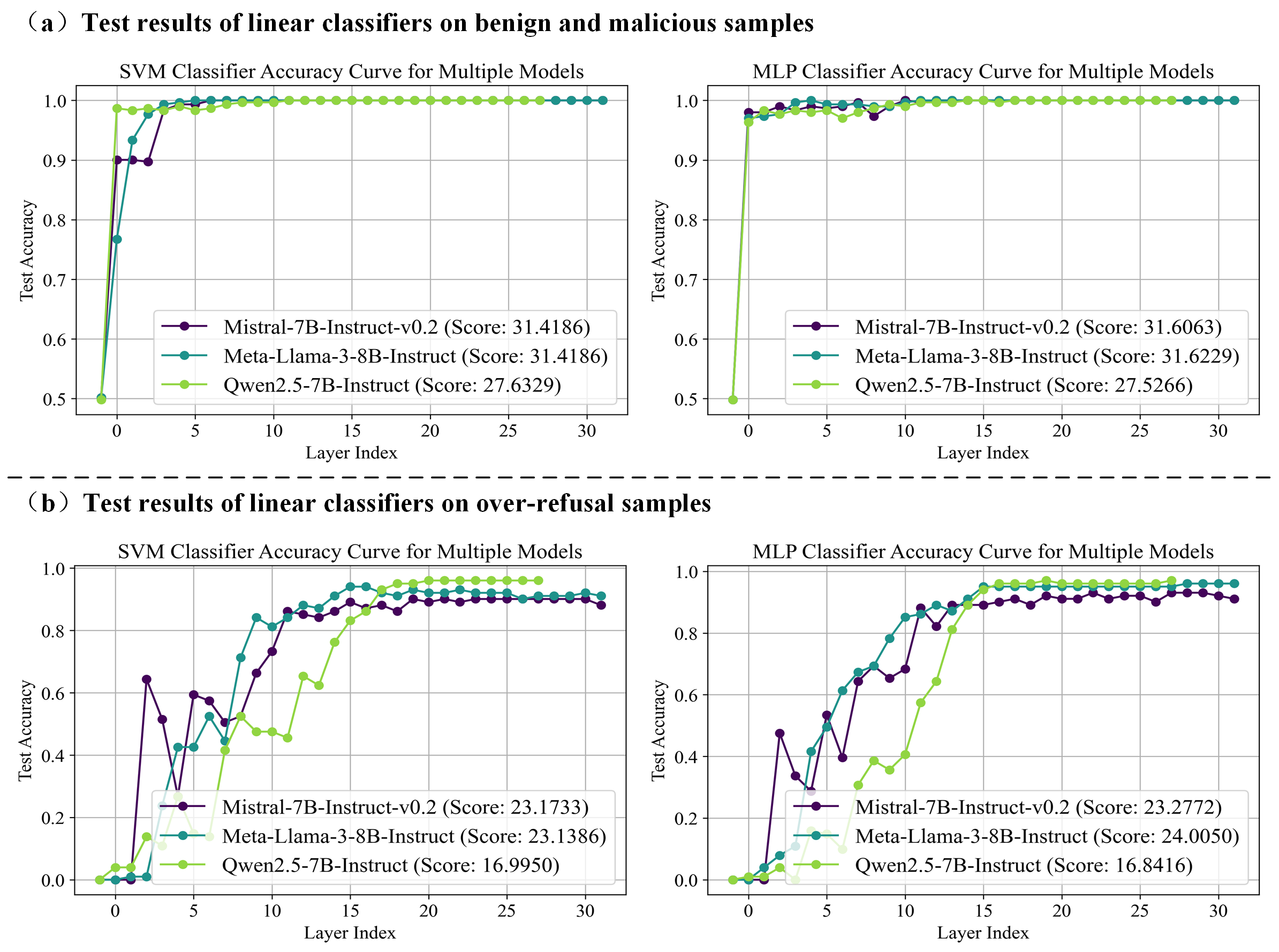}
	\caption{Probe classification results with 1000 samples per category.}
	\label{fig: prob_1000}
\end{figure}

In the probe experiments, we expand the sample size for each class to 1000 and repeat the validation. 
Fig. \ref{fig: prob_1000} shows the experimental results with 1,000 samples, which are fully consistent with the conclusions drawn from the original 100-sample set. 
Across different sample sizes, weak classifiers consistently maintain high classification accuracy on normal and malicious samples, while accuracy remains low on over-refusal samples. 
This indicates that only a small number of samples is sufficient to effectively capture the security representation concepts within large models.

\subsection{Safety Layer}

\begin{figure}[H]
	\centering
	\includegraphics[width=0.48\textwidth]{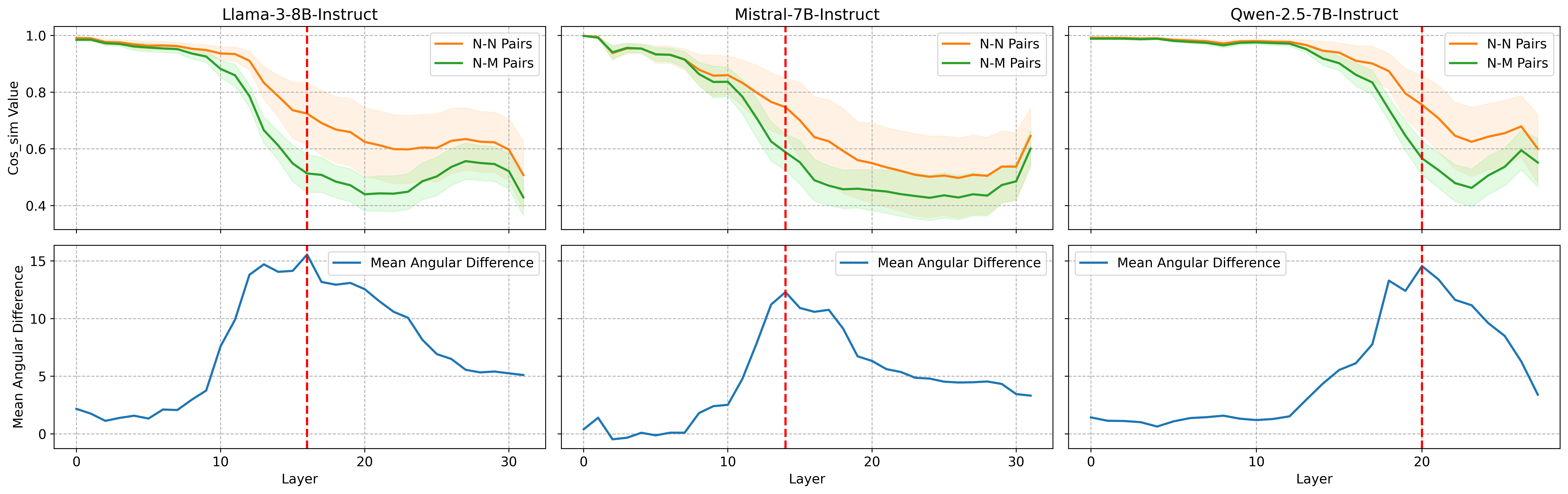}
	\caption{Layer-wise safety representation analysis via cosine similarity and angular difference.}
	\label{fig:safetylayer}
\end{figure}

In the probe analysis experiments, the weak classifier achieves high classification accuracy on some layers of the model, while exhibiting significantly low accuracy on other layers. 
Recent studies\cite{li2025safety} have revealed safety layers within LLMs, suggesting that analyses of probe classifier accuracy should focus on specific layers. 
To further investigate the sensitivity of safety layers, we designed and conducted a layer-wise cosine similarity analysis experiment.
This experiment systematically compares the hidden-layer feature similarities across three mainstream models: Llama3-8B-Instruct, Mistral-7B-Instruct-v0.2, and Qwen2.5-7B-Instruct. The experimental procedure is as follows: First, we construct two types of text pairs—normal–normal (N-N) and normal–malicious (N-M)—through random sampling. 
To ensure statistical robustness, each type of text pair is independently sampled 500 times. Next, we feed the text into the models and extract the hidden vectors of the last token in each layer, then compute the cosine similarity between the representations of the two groups layer by layer to obtain curves of average similarity against model depth. 
Finally, we quantify representational divergence across layers by averaging the angular difference between the two similarity curves. Normal samples are drawn from the Alpaca dataset, and malicious samples use harmful prompts from jailbreak attacks.

Fig. \ref{fig:safetylayer} presents the key results of this experiment. 
First, from the trend of the individual mean curves: comparing the layer-wise cosine similarity of normal–normal (N-N) and normal–malicious (N-M) query pairs, the N-M curve shows a significant divergence, which directly validates the existence of safety layers. 
The representational differences across layers when the model distinguishes normal and malicious queries are far larger than those between normal queries with different semantics, and this trend is consistent across all aligned LLMs. 
Second, from the average angular difference curves, the curves are smooth, and the angular difference is nearly zero in the earliest layers, and the representations of the two query types highly overlap. 
This indicates that safety-related features are not yet activated at this stage; the model processes malicious and normal queries identically, and the lower layers cannot identify malicious content. 
Starting from several middle layers, the gap between the curves begins to widen rapidly, rises steeply, and then stabilizes. 
This significant expansion of the gap marks that the aligned large language model officially activates the mechanism to distinguish normal and malicious queries, serving as direct evidence of the functioning safety layers.
In summary, in probe experiments, we should focus on the classifier performance on safety layers. Low classifier accuracy on safety layers indicates that the model struggles to effectively distinguish normal samples from over-refusal samples.

\subsection{In-Depth Analysis of Context-Aware Enhancement}
\begin{figure}[h]
	\centering
	\includegraphics[width=0.35\textwidth]{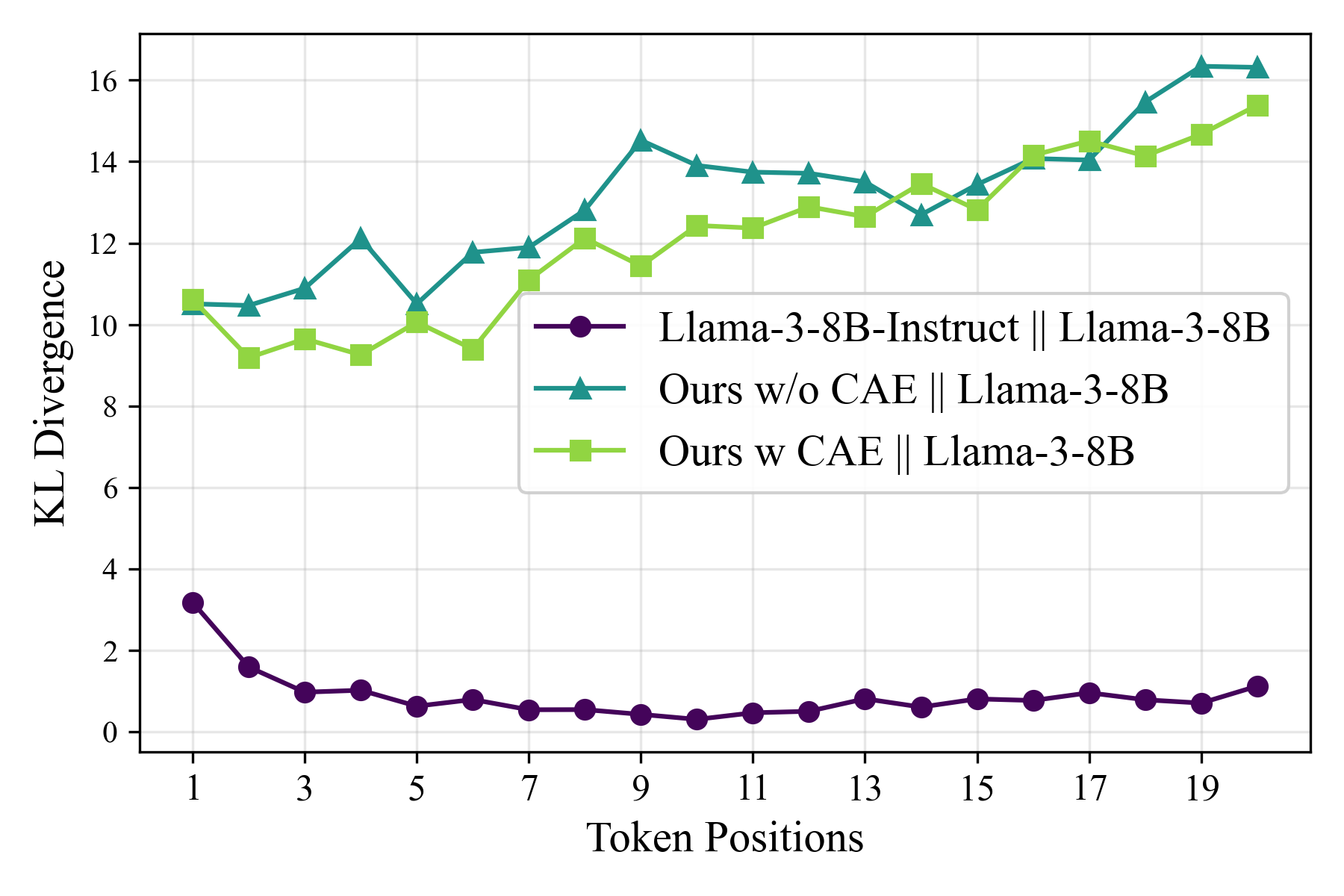}
	\caption{KL divergence difference comparison.}
	\label{fig9}
\end{figure}

In this subsection, we explore the fundamental reasons behind the effectiveness of context-aware enhancement (CAE).
For standard safety-aligned data, immediately following a harmful prompt with a rejection response directly overcomes adversarial examples.
In this manner, safety alignment essentially increases the probability of the rejection token at the initial position of the response.
However, building upon our previous experiments on the causes of model over-refusal, relying solely on the initial prompt leads the model to misclassify benign samples as adversarial ones.
Therefore, we aim to weaken the probability of the initial token being a rejection token.
To demonstrate this, we conduct an experiment.
Specifically, we collect adversarial prompts from the Harmbench dataset, where all responses were rejection replies.
We compute the per-token KL divergence for harmful prompts before and after model alignment. 
Fig. \ref{fig9} shows the results for the first 20 tokens.

First, for the safety-aligned version Llama3-8B-Instruct, the KL divergence budget for safety is primarily concentrated in the initial few tokens compared to the pre-trained Llama3-8B model.
Second, without context-aware augmentation, the KL divergence differences per token become more pronounced, leading to deeper alignment by defense methods.
This highlights the trade-off between deep alignment and practicality.
Finally, when context-aware augmentation is applied, the KL divergence differences per token are mitigated.
At a high level, achieving better safety requires deeper alignment and stronger alignment.
To account for over-refusal, we need to adjust alignment strength—a straightforward adjustment at the data level.

Tables \ref{tab3} and \ref{tab4} show the performance for different harmful prefix lengths.
When $\mathcal{L}=10$, the model achieves optimal balance; excessively long, harmful prefixes prevent the model from redirecting to rejection responses.

\begin{table}[H]
	\caption{Impact of the hyperparameter context length $\mathcal{L}$ on Llama3-8B-Instruct.}
	\label{tab3}
	\centering
	\resizebox{\columnwidth}{!}{
		\begin{tabular}{cccccccc} 
			\toprule[1.5pt]
			\multirow{2}{*}{\textbf{Models}} & \multirow{2}{*}{\textbf{Length}} & \multicolumn{2}{c}{\textbf{Safety ASR (\%) $\downarrow$}} & \multicolumn{3}{c}{\textbf{Over-Refusal Rate (\%) $\downarrow$}} & \multirow{2}{*}{\textbf{Tradeoff Score}} \\
			\cmidrule(lr){3-4} \cmidrule(lr){5-7} 
			& & WildGuardTest & HarmBench & PHTest & OR-Bench-1K & FalseReject & \\
			\cmidrule(lr){1-2} \cmidrule(lr){3-4} \cmidrule(lr){5-7} \cmidrule(lr){8-8}
			% \midrule 
			\multirow{5}{*}{\textbf{Llama3-8B-Instruct}} 
			& $\mathcal{L}=0$  &6.37  &11.67  &17.67  &74.67  &47.35  & 31.55 \\
			& $\mathcal{L}=10$ &8.35  &15.42  &14.67  &67.78  &36.90  & \textbf{28.62} \\
			& $\mathcal{L}=20$ & 8.75 &15.00  &15.00  &66.26  &40.01  & 29.00 \\
			& $\mathcal{L}=30$ & 9.68 &16.67  &13.00  &66.41  &39.60  & 29.07 \\
			& $\mathcal{L}=40$ & 6.89 &15.83  &15.00  &70.73  &42.46  & 30.18 \\
			\bottomrule[1.5pt]
		\end{tabular}
	}	
\end{table}

\vspace{-5mm}

\begin{table}[H]
	\caption{Impact of the hyperparameter context length $\mathcal{L}$ on Mistral-7B-Instruct-v0.2.}
	\label{tab4}
	\centering
	\resizebox{\columnwidth}{!}{
		\begin{tabular}{cccccccc} 
			\toprule[1.5pt]
			\multirow{2}{*}{\textbf{Models}} & \multirow{2}{*}{\textbf{Length}} & \multicolumn{2}{c}{\textbf{Safety ASR (\%) $\downarrow$}} & \multicolumn{3}{c}{\textbf{Over-Refusal Rate (\%) $\downarrow$}} & \multirow{2}{*}{\textbf{Tradeoff Score}} \\
			\cmidrule(lr){3-4} \cmidrule(lr){5-7} % 使用 cmidrule 并修剪两端
			& & WildGuardTest & HarmBench & PHTest & OR-Bench-1K & FalseReject & \\
			\cmidrule(lr){1-2} \cmidrule(lr){3-4} \cmidrule(lr){5-7} \cmidrule(lr){8-8}
			\multirow{5}{*}{\textbf{Mistral-7B-Instruct-v0.2}} 
			& $\mathcal{L}=0$ & 4.24 & 11.25 &19.33  &49.96  &40.35  & 25.03 \\
			& $\mathcal{L}=10$ & 10.74 &20.42  &15.00  &27.14  & 23.10 & \textbf{19.28} \\
			& $\mathcal{L}=20$ & 23.87 &32.08  &13.00  &23.28  & 23.17 & 23.08 \\
			& $\mathcal{L}=30$ & 28.65 &46.67  &13.67  &17.73  &22.16  & 25.78 \\
			& $\mathcal{L}=40$ &17.11  &22.50  &13.33  & 25.40 &24.52  & 20.57 \\
			\bottomrule[1.5pt]
		\end{tabular}
	}
\end{table}

\vspace{-32pt}

\subsection{Empirical Analysis on Cosine Similarity}

\begin{figure}[H]
	\centering
	\includegraphics[width=0.48\textwidth]{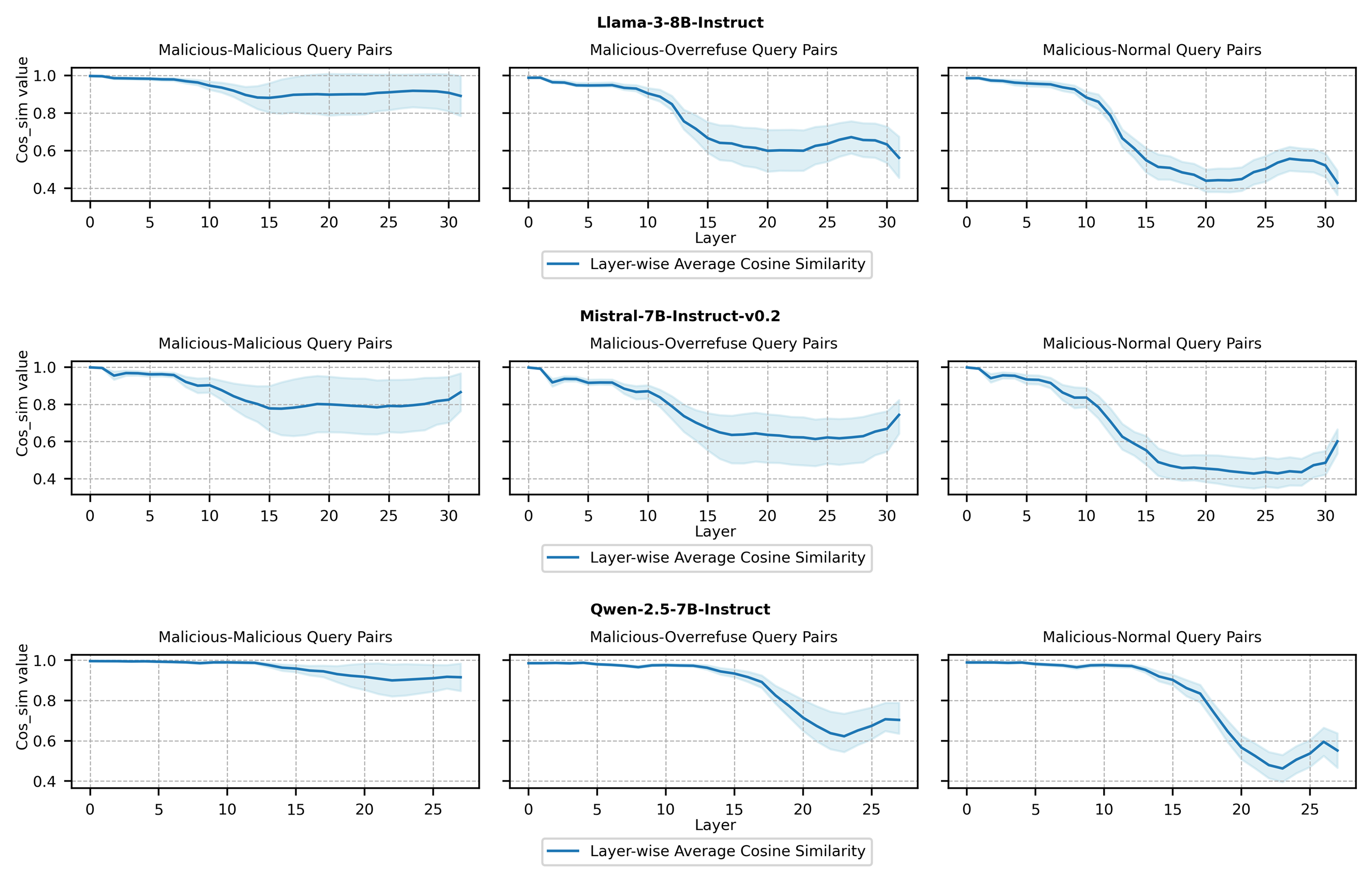}
	\caption{Layer-wise cosine similarity analysis when processing different types of prompts on three models.}
	\label{fig:cossim}
\end{figure}

We define overlap directly as the cosine similarity between sample representations, which quantifies how close two samples lie in the representation space. 
According to the theory of orthogonal gradient descent\cite{farajtabar2020orthogonal}, if an over-refusal sample and a malicious sample exhibit high cosine similarity, they will share similar gradient descent directions. 
Similar observations have been reported in studies on safety degradation induced by benign fine-tuning\cite{he2024what} and in multilingual safety research\cite{he2025tuba}. 
We conduct further quantitative experiments: specifically, we compute the layer-wise cosine similarity changes for three types of text pairs, namely malicious–malicious (M-M), malicious–overrefusal (M-O), and malicious–normal (M-N), following a similar analysis to that in the preceding section.
In non-security layers (such as shallow layers), the inter-layer differences among these sample categories are relatively consistent. 
Fig. \ref{fig:cossim} gives the analysis results.
Compared to normal samples, over-refusal samples and malicious samples exhibit higher cosine similarity.

\subsection{Ablation Study}

We construct ablation experiments on both the overlap-aware loss weighting and context-aware augmentation components.
Table \ref{tab6} and Table \ref{tab7} present experimental results across two models.
First, without considering over-refusal mitigation, the defense-enhanced model baseline exhibits lower ASR but higher over-refusal rates.
Second, overlap-aware loss weighting achieves better results in mitigating over-refusal, demonstrating the effectiveness of intervening in the loss weighting of malicious samples during training.
Context-aware augmentation also reduces over-refusal rates, but it results in decreased safety; however, the balanced score remains superior to the baseline.
This aligns with our vision that enhancing model safety should not come at the sole expense of answering benign queries; we should strive for a balanced approach.
Finally, the balanced score achieves the best performance when both overlap-aware loss weighting and context-aware augmentation are applied simultaneously.
Additionally, we observe that mitigation is more effective for models with initially higher safety (e.g., Llama), akin to adjusting a spring from an unbalanced, overly taut state to a more balanced position.

\begin{table}[H]
	\caption{Ablation study on Llama3-8B-Instruct. In this table, baseline represents the scenario without over-refusal mitigation; A represents overlap-aware loss weighting; B represents context-aware enhancement, and C represents the simultaneous use of both.}
	\label{tab6}
	\centering
	\resizebox{\columnwidth}{!}{
		\begin{tabular}{c|cc|cccccc}
			\midrule
			\multirow{2.5}{*}{\textbf{Models}} & \multirow{2.5}{*}{A} & \multirow{2.5}{*}{B} & \multicolumn{2}{c}{\textbf{Safety ASR (\%) $\downarrow$}} & \multicolumn{3}{c}{\textbf{Over-Refusal Rate (\%) $\downarrow$}} & \multirow{2.5}{*}{\textbf{Tradeoff Score}} \\
			\cmidrule(lr){4-5} \cmidrule(lr){6-8} 
			~ & & & WildGuardTest & HarmBench & PHTest & OR-Bench-1K & FalseReject & \\
			\midrule
			Baseline & & &6.37  &11.67  &17.67  &74.67  &47.35  &31.55  \\
			\cmidrule(lr){1-9} 
			(a) & $\checkmark$ & & 12.86 &9.58  &10.67  &54.59  &31.43  &23.83  \\
			(b) &  & $\checkmark$ &8.35  &15.42  &14.67  &67.78  &36.90  &28.62  \\
			(c) & $\checkmark$ & $\checkmark$ & 8.21 & 11.25 &11.00  &41.70  &26.12  &19.66  \\
			\bottomrule[1.5pt]
	\end{tabular}}
\end{table}

\vspace{-5mm}

\begin{table}[H]
	\caption{Ablation study on Mistral-7B-Instruct-v0.2.}
	\label{tab7}
	\centering
	\resizebox{\columnwidth}{!}{
		\begin{tabular}{c|cc|cccccc}
			\toprule[1.5pt]
			\multirow{2}{*}{\textbf{Models}} & 
			\multirow{2}{*}{A} & 
			\multirow{2}{*}{B} & 
			\multicolumn{2}{c}{\textbf{Safety ASR (\%) $\downarrow$}} & 
			\multicolumn{3}{c}{\textbf{Over-Refusal Rate (\%) $\downarrow$}} & 
			\multirow{2}{*}{\textbf{Tradeoff Score}} \\
			\cmidrule(lr){4-5} \cmidrule(lr){6-8} 
			~ & & & WildGuardTest & HarmBench & PHTest & OR-Bench-1K & FalseReject & \\
			\midrule
			Baseline & & &4.24  &11.25  &19.33  &49.96  &40.35  &25.03  \\
			\cmidrule(lr){1-9} 
			(a) & $\checkmark$ & &9.55  &26.25  &13.33  &30.40  &22.32  &20.37  \\
			(b) &  & $\checkmark$ &9.68  &17.92  &15.00  &35.63  &31.00  &21.85 \\
			(c) & $\checkmark$ & $\checkmark$ &13.66  &22.71  &6.83  &27.94  &28.35  &19.89  \\
			\bottomrule[1.5pt]
		\end{tabular}
	}
\end{table}

We initially observed that directly combining overlap-aware loss weighting and context-aware enhancement using the hyperparameters tuned for each individual mechanism did not achieve satisfactory performance on Mistral-7B-Instruct-v0.2. 
To understand this phenomenon, we provide a heuristic explanation using a spring analogy.

An instruction-tuned model can be viewed as a safety-aligned system with a certain baseline refusal tendency, while overlap-aware loss weighting and context-aware enhancement can be regarded as relaxation operations applied to this system, aiming to reduce the over-refusal behavior caused by overly conservative safety alignment.

For Llama3-8B-Instruct, the model exhibits a stronger baseline refusal tendency, which can be analogized to a spring under relatively high tension. 
In this case, only a small amount of “release” is required to induce noticeable behavioral changes. 
Correspondingly, when the two mechanisms are applied together, the combined approach is more likely to achieve an improved safety–helpfulness trade-off even without extensive additional hyperparameter tuning.

In contrast, Mistral-7B-Instruct-v0.2 exhibits a relatively weaker initial refusal tendency, analogous to a spring that is already in a more relaxed state. 
In this case, simply stacking two mechanisms designed to mitigate over-refusal may result in insufficient adjustment or a shift away from the desired direction. 
The model requires more careful “elastic adjustment” to move toward an appropriate safety–helpfulness balance. 
Therefore, the joint effectiveness of overlap-aware loss weighting and context-aware enhancement is more sensitive to training hyperparameters, requiring further task-specific calibration and optimization.

\subsection{Temperature coefficient}

In the overlap-aware loss weighting, the temperature parameter controls the sharpness of the sample weight distribution within each batch.
To investigate its impact, we conducted experiments using a set of temperature values: {0.1, 0.5, 1, 2, 4}.
Quantitative results are summarized in Table \ref{tab8} and Table \ref{tab9}. 
The Llama and Mistral models achieved optimal balanced scores when $\tau=1$ and $\tau=4$, respectively.

\begin{table}[H]
	\caption{Impact of the hyperparameter temperature $\tau$ on Llama3-8B-Instruct.}
	\label{tab8}
	\centering
	\resizebox{\columnwidth}{!}{
		\begin{tabular}{cccccccc} % 移除了所有竖线
			\toprule[1.5pt]
			\multirow{2}{*}{\textbf{Models}} & \multirow{2}{*}{\textbf{Temperature}} & \multicolumn{2}{c}{\textbf{Safety ASR (\%) $\downarrow$}} & \multicolumn{3}{c}{\textbf{Over-Refusal Rate (\%) $\downarrow$}} & \multirow{2}{*}{\textbf{Tradeoff Score}} \\
			\cmidrule(lr){3-4} \cmidrule(lr){5-7} % 使用 cmidrule 并修剪两端
			& & WildGuardTest & HarmBench & PHTest & OR-Bench-1K & FalseReject & \\
			\cmidrule(lr){1-2} \cmidrule(lr){3-4} \cmidrule(lr){5-7} \cmidrule(lr){8-8}
			\multirow{6}{*}{\textbf{Llama3-8B-Instruct}} 
			& Vanilla & 6.37 &11.67  &17.67  &74.67  &47.35  & 31.55 \\
			& $\tau=0.1$ & 3.18 &3.33  &21.33  &78.08  &44.65  & 30.11 \\
			& $\tau=0.5$ & 6.10 & 5.00 & 16.00 &63.30  &41.01  & 26.28 \\
			& $\tau=1$ & 6.23 &6.67  &12.33  &58.15  &35.30  & \textbf{23.74} \\
			& $\tau=2$ &12.86  &9.58  &10.67  &54.59  &31.43  & 23.83 \\
			& $\tau=4$ & 4.11 & 2.92 & 18.33 &70.35  &44.06  & 27.95 \\
			\bottomrule[1.5pt]
		\end{tabular}
	}
\end{table}

\vspace{-5mm}

\begin{table}[H]
	\caption{Impact of the hyperparameter temperature $\tau$ on Mistral-7B-Instruct-v0.2.}
	\label{tab9}
	\centering
	\resizebox{\columnwidth}{!}{
		\begin{tabular}{cccccccc} 
			\toprule[1.5pt]
			\multirow{2}{*}{\textbf{Models}} & \multirow{2}{*}{\textbf{Temperature}} & \multicolumn{2}{c}{\textbf{Safety ASR (\%) $\downarrow$}} & \multicolumn{3}{c}{\textbf{Over-Refusal Rate (\%) $\downarrow$}} & \multirow{2}{*}{\textbf{Tradeoff Score}} \\
			\cmidrule(lr){3-4} \cmidrule(lr){5-7} 
			& & WildGuardTest & HarmBench & PHTest & OR-Bench-1K & FalseReject & \\
			\cmidrule(lr){1-2} \cmidrule(lr){3-4} \cmidrule(lr){5-7} \cmidrule(lr){8-8}
			\multirow{6}{*}{\textbf{Mistral-7B-Instruct-v0.2}} 
			& Vanilla &4.24  &11.25  &19.33  &49.96  &40.35  & 25.03 \\
			& $\tau=0.1$ &10.34  &16.67  &15.33  &35.55  &32.27  & 22.03 \\
			& $\tau=0.5$ &10.61  &17.08  &17.67  &44.66  &32.94  & 24.59 \\
			& $\tau=1$ & 18.30 &23.75  &17.33  &32.37  &27.13  & 23.78 \\
			& $\tau=2$ &23.87  &34.58  &15.00  &20.85  &25.19  & 23.90 \\
			& $\tau=4$ & 9.55 &26.25  &13.33  &30.40  &22.32  & \textbf{20.37} \\
			\bottomrule[1.5pt]
		\end{tabular}
	}
	
\end{table}

\subsection{Utility Evaluation}

\begin{table}[H]
	\caption{Utility evaluation results for Mistral-7B-Instruct-v0.2. A higher value is better.}
	\label{tab:mistral}
	\centering
	\resizebox{\columnwidth}{!}{
		\begin{tabular}{lcccccccc}
			\toprule[1.5pt]
			\textbf{Benchmark} & \textbf{Vanilla} & \textbf{TA} & \textbf{NPO} & \textbf{DPO} & \textbf{WHP} & \textbf{CB} & \textbf{RB} & \textbf{Ours} \\
			\cmidrule(lr){1-9}
MMLU & 59.06$\pm$0.39 & 58.15$\pm$0.39 & 58.68$\pm$0.40 & 58.64$\pm$0.39 & 58.87$\pm$0.39 & 59.02$\pm$0.39 & 58.99$\pm$0.39 & 58.62$\pm$0.39 \\
GSM8K & 42.61$\pm$1.36 & 42.53$\pm$1.36 & 39.35$\pm$1.35 & 42.91$\pm$1.36 & 42.53$\pm$1.36 & 42.23$\pm$1.36 & 41.55$\pm$1.36 & 42.84$\pm$1.36 \\
TruthfulQA & 55.32$\pm$1.74 & 58.26$\pm$1.73 & 53.12$\pm$1.75 & 52.88$\pm$1.75 & 53.37$\pm$1.75 & 53.61$\pm$1.75 & 59.61$\pm$1.72 & 54.35$\pm$1.74 \\
Winogrande & 73.80$\pm$1.24 & 73.16$\pm$1.25 & 73.16$\pm$1.25 & 74.11$\pm$1.23 & 70.80$\pm$1.28 & 73.40$\pm$1.24 & 74.27$\pm$1.23 & 72.53$\pm$1.25 \\
ARC & 81.27$\pm$0.80 & 81.19$\pm$0.80 & 82.95$\pm$0.77 & 79.42$\pm$0.83 & 77.95$\pm$0.85 & 81.40$\pm$0.80 & 80.93$\pm$0.81 & 81.61$\pm$0.79 \\
HellaSwag & 83.73$\pm$0.37 & 79.96$\pm$0.40 & 84.22$\pm$0.36 & 83.96$\pm$0.37 & 81.19$\pm$0.39 & 83.50$\pm$0.37 & 83.06$\pm$0.37 & 81.30$\pm$0.39 \\
			\bottomrule[1.5pt]
		\end{tabular}
	}
\end{table}

\vspace{-5mm}

\begin{table}[H]
	\caption{Utility evaluation results for Llama3-8B-Instruct. A higher value is better.}
	\label{tab:llama}
	\centering
	\resizebox{\columnwidth}{!}{
		\begin{tabular}{lcccccccc}
			\toprule[1.5pt]
			\textbf{Benchmark} & \textbf{Vanilla} & \textbf{TA} & \textbf{NPO} & \textbf{DPO} & \textbf{WHP} & \textbf{CB} & \textbf{RB} & \textbf{Ours} \\
			\cmidrule(lr){1-9}
		MMLU & 63.81$\pm$0.38 & 62.76$\pm$0.39 & 63.42$\pm$0.38 & 63.28$\pm$0.38 & 63.91$\pm$0.38 & 63.75$\pm$0.38 & 63.69$\pm$0.38 & 63.69$\pm$0.38 \\
		GSM8K & 75.82$\pm$1.18 & 71.49$\pm$1.24 & 73.46$\pm$1.22 & 76.27$\pm$1.17 & 75.82$\pm$1.18 & 75.66$\pm$1.18 & 75.06$\pm$1.19 & 75.79$\pm$1.18 \\
		TruthfulQA & 47.25$\pm$1.75 & 48.47$\pm$1.75 & 47.25$\pm$1.75 & 43.33$\pm$1.73 & 49.82$\pm$1.75 & 48.47$\pm$1.75 & 51.77$\pm$1.75 & 47.11$\pm$1.75 \\
		Winogrande & 71.43$\pm$1.27 & 72.69$\pm$1.25 & 72.30$\pm$1.26 & 71.67$\pm$1.27 & 65.04$\pm$1.34 & 71.51$\pm$1.27 & 72.22$\pm$1.26 & 71.16$\pm$1.26 \\
		ARC & 81.65$\pm$0.79 & 81.78$\pm$0.79 & 81.99$\pm$0.79 & 80.77$\pm$0.81 & 78.32$\pm$0.85 & 81.40$\pm$0.80 & 81.44$\pm$0.80 & 81.14$\pm$0.80 \\
		HellaSwag & 75.81$\pm$0.43 & 76.14$\pm$0.43 & 76.20$\pm$0.42 & 75.60$\pm$0.43 & 69.03$\pm$0.46 & 75.87$\pm$0.43 & 74.89$\pm$0.43 & 75.85$\pm$0.43 \\
			
			\bottomrule[1.5pt]
		\end{tabular}
	}
\end{table}

\begin{figure}[H]
	\centering
	\includegraphics[width=0.49\textwidth]{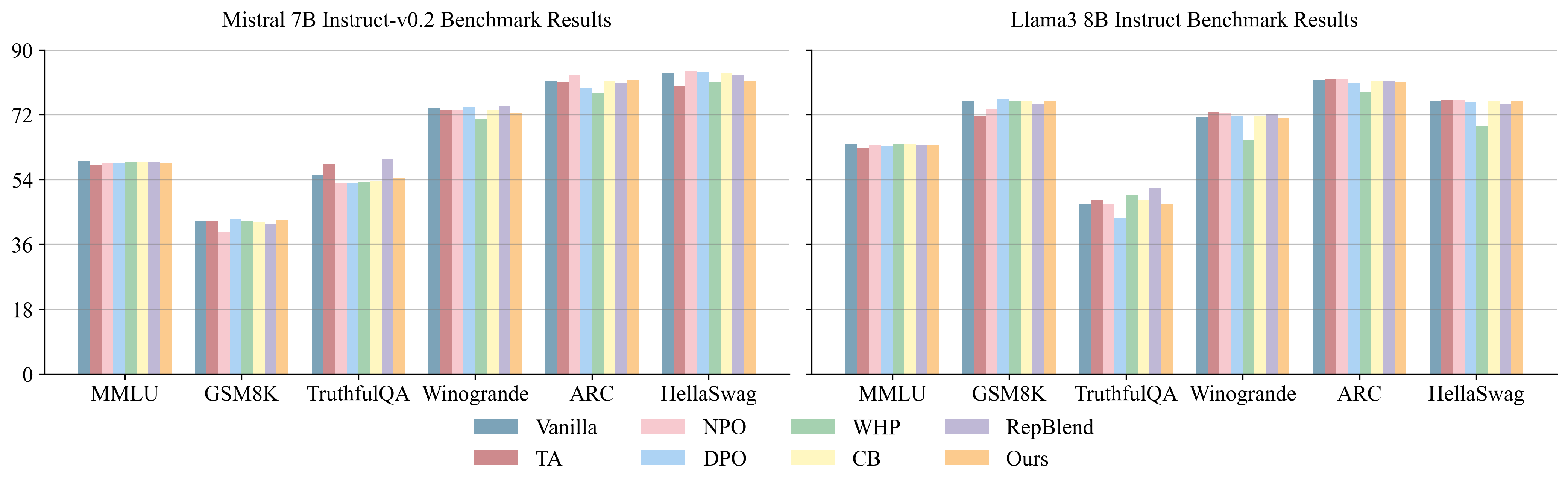}
	\caption{Utility evaluation results.}
	\label{fig:Utility}
\end{figure}

We repeated the utility evaluation in three independent runs, and the results of all tasks are reported as mean \(\pm\) standard deviation.
Tables \ref{tab:llama} and \ref{tab:mistral} show the corresponding experimental results. 
Compared to the original model, the decline in utility caused by our method does not exceed 1.5\%, indicating that it effectively preserves the original model’s general capabilities. 
The overall utility is roughly equivalent to that of the baseline model, suggesting that safety alignment and the mitigation of over-refusal do not significantly impair the model’s foundational capabilities.
Fig. \ref{fig:Utility} provides a visual comparison of the results.

\subsection{Training Cost}
	Table \ref{tab:efficient} compares the training time and GPU memory consumption of DPO, NPO, WHP, TA, RB, and our proposed method on the Llama3-8B-Instruct model with NVIDIA L40 GPUs. 
	Our approach obtains the minimal training time and memory overhead, showing obvious advantages in computational efficiency.

\begin{table}[H]
	\centering
	\caption{Training cost evaluation results on Llama3-8B-Instruct with NVIDIA L40 GPU.}
	\label{tab:efficient}
	\small
	\begin{tabular}{l c c}
		\hline
		\textbf{Method} & \textbf{Time} & \textbf{Memory} \\
		\hline
		DPO   & 3.5 Hour  & 38.8GB \\
		NPO   & 7.5 Hours & 37.5GB \\
		WHP   & 4 Hours & 43.2GB \\
		TA    & 1.5 Hours & 30GB \\
		RB    & 5.2 Hour & 44.8GB \\
		Ours  & \textbf{1 Hour}  & \textbf{30GB} \\
		\hline
	\end{tabular}
\end{table}

\section{Conclusion}
In this paper, we treat model representations as fundamental units of analysis, conducting a multi-level investigation into why benign samples are incorrectly rejected. Ultimately, we attribute this to the fact that these over-refusal representations lie in the boundary region between malicious and benign samples.
Based on this, we propose two strategies to mitigate over-refusal during LLM alignment: (1) constraining the rejection training process by reducing the contribution of malicious samples similar to over-refusal samples to the model's rejection capability, and (2) employing simple context-aware data augmentation.
Experimental results demonstrate that our approach mitigates the tendency for over-refusal sample representations to migrate toward the rejection representation space during the alignment process.
Fundamentally, due to the high similarity between over-refusal and malicious samples, achieving both high rejection of adversarial samples and low over-refusal in LLM alignment remains challenging, as demonstrated by extensive comparative experiments.
We advocate that future defenses should better balance safety and over-refusal rather than solely pursuing alignment depth.

In the future, we will explore the following directions:
(1) Multi‐turn over-refusal scenarios. 
This paper focuses on single-round scenarios; however, no benchmarks have been developed explicitly for multi-round over-refusal scenarios to date. Assessing multi-round over-refusal, multi-round safety, and their interrelationship constitutes an essential direction for future research.
(2) Lower-cost mitigation approaches.
Fine-tuning LLMs remains costly, and we will explore dynamic reasoning intervention methods based on representation engineering in the future.

% {\appendix[Proof of the Zonklar Equations]
% Use $\backslash${\tt{appendix}} if you have a single appendix:
% Do not use $\backslash${\tt{section}} anymore after $\backslash${\tt{appendix}}, only $\backslash${\tt{section*}}.
% If you have multiple appendixes use $\backslash${\tt{appendices}} then use $\backslash${\tt{section}} to start each appendix.
% You must declare a $\backslash${\tt{section}} before using any $\backslash${\tt{subsection}} or using $\backslash${\tt{label}} ($\backslash${\tt{appendices}} by itself
%  starts a section numbered zero.)}


% \section{References Section}
% You can use a bibliography generated by BibTeX as a .bbl file.
%  BibTeX documentation can be easily obtained at:
%  http://mirror.ctan.org/biblio/bibtex/contrib/doc/
%  The IEEEtran BibTeX style support page is:
%  http://www.michaelshell.org/tex/ieeetran/bibtex/
 
\bibliographystyle{IEEEtran}
\normalem
\bibliography{manuscript}

\newpage

 

% \end{IEEEbiography}



\vfill

\end{document}